\documentclass[aps,amsmath,amssymb,physrev,reprint,superscriptaddress]{revtex4-2}

\usepackage{graphicx}
\usepackage{dcolumn}
\usepackage{bm}
\usepackage[table,xcdraw]{xcolor}
\usepackage{amsmath}
\usepackage[utf8]{inputenc}
\usepackage{siunitx}
\usepackage{graphicx}
\usepackage[normalem]{ulem}
\usepackage{xcolor}
\usepackage{tabularx}
\usepackage{siunitx}
\usepackage{textcomp}
\usepackage{gensymb}
\usepackage{multirow}
\usepackage{color, colortbl}

\definecolor{Gray}{gray}{0.9}
\definecolor{LightCyan}{rgb}{0.88,1,1}
\definecolor{Red}{cmyk}{0, 0.7808, 0.4429, 0.1412}

\makeatletter
\def\CT@@do@color{%
  \global\let\CT@do@color\relax
        \@tempdima\wd\z@
        \advance\@tempdima\@tempdimb
        \advance\@tempdima\@tempdimc
\advance\@tempdimb\tabcolsep
\advance\@tempdimc\tabcolsep
\advance\@tempdima2\tabcolsep
        \kern-\@tempdimb
        \leaders\vrule
                \hskip\@tempdima\@plus  1fill
        \kern-\@tempdimc
        \hskip-\wd\z@ \@plus -1fill }
\makeatother

\begin{document}

\preprint{APS/123-QED}

\title{Ultra-low Threshold Titanium doped sapphire Whispering-gallery Laser}
\author{Farhan Azeem}
\author{Luke S. Trainor} 

\affiliation{The Dodd-Walls Centre for Photonic and Quantum Technologies, New Zealand}
\affiliation{Department of Physics, University of Otago, 730 Cumberland Street, Dunedin 9016, New Zealand}

\author{Ang Gao}
\affiliation{MOE Key Laboratory of Weak-Light Nonlinear Photonics, TEDA Institute of Applied Physics and School of Physics, Nankai University, Tianjin, 300457, China}
\affiliation{State Key Laboratory for Artificial Microstructure and Mesoscopic Physics, School of Physics, Peking University, Beijing, 100871, China}

\author{Maya Isarov}
\author{Dmitry V. Strekalov}
\author{Harald G. L. Schwefel}
\email{harald.schwefel@otago.ac.nz}
\affiliation{The Dodd-Walls Centre for Photonic and Quantum Technologies, New Zealand}
\affiliation{Department of Physics, University of Otago, 730 Cumberland Street, Dunedin 9016, New Zealand}

\date{\today}

\begin{abstract}
Titanium doped sapphire (Ti:sapphire) is a laser gain material with broad gain bandwidth benefiting from the material stability of sapphire. These favorable characteristics of Ti:sapphire have given rise to femtosecond lasers and optical frequency combs. Shaping a single Ti:sapphire crystal into a millimeter sized high quality whispering gallery mode resonator ($Q\sim10^8$) reduces the lasing threshold to \SI{14.2}{\milli\W} and increases the laser slope efficiency to 34\%. The observed lasing can be both multi-mode and single-mode. This is the first demonstration of a Ti:sapphire whispering gallery laser. 
Furthermore, a novel method of evaluating the gain in Ti:sapphire in the near infrared region is demonstrated by introducing a probe laser with a central wavelength of \SI{795}{\nano\m}. This method results in decreasing linewidth of the modes excited with the probe laser, consequently increasing their $Q$. These findings open avenues for the usage of whispering gallery mode resonators as cavities for the implementation of compact Ti:sapphire lasers. Moreover, Ti:sapphire can also be utilized as an amplifier inside its gain bandwidth by implementing a pump-probe configuration.

\end{abstract}

                           
\maketitle

\section{\label{sec:level1} Introduction}

Titanium doped sapphire (Ti:sapphire) has been the workhorse behind solid state lasers since its invention in 1986~\cite{moulton1986spectroscopic}. Its broad gain bandwidth -- ranging from \SI{650}{\nano\m} up to \SI{1100}{\nano\m} -- and peak absorption wavelengths between \SI{514}{\nano\m} and \SI{532}{\nano\m}~\cite{moulton1992tunable} made it the gain material of choice for solid state lasers~\cite{spence199160}, ultra-short pulse lasers~\cite{gibson1996electro}, wide wavelength range tunable lasers, and femtosecond lasers~\cite{bartels2008passively,gurel2015green}. Adding on top sapphire's excellent thermal conductivity and material strength, Ti:sapphire oscillators are used to generate femtosecond pulses with megahertz repetition rates~\cite{naumov2005approaching,kalashnikov2005approaching} and high-power chirped pulse amplification \cite{kiriyama2009generation}, generating output powers as high as \SI{4.2}{\peta\watt}~\cite{sung20174}. Moreover, Ti:sapphire lasers have also been used to demonstrate frequency combs~\cite{matos2004direct} and to study ultrafast dynamics of atomic, molecular, and condensed matter~\cite{krausz2009attosecond,frietsch2013high,chiang2012high}.

Here we present a new take on the material that will allow us to surpass the current records in lasing threshold and slope efficiency by fabricating an ultra-high quality whispering gallery mode resonator out of Ti:Sapphire. The lasing threshold is directly reliant on the quality factor ($Q$-factor) of the resonator and the volume ($V$) of the mode inside the cavity. The $Q$-factor gives a measure of the optical loss inside the cavity, whereas $V$ affects the field strength inside the cavity~\cite{he2013whispering} via the Purcell effect. A high $Q$-factor and a low $V$, give rise to stronger coupling between gain medium and the field of the cavity~\cite{pelton2002three,kimble1998strong,ozdemir2011estimation}. This in turn translates into a possible low lasing threshold and narrow linewidths~\cite{bulovic1998transform,spiegelberg2004low}. 
In lasers the gain compensates the losses, therefore slight improvements in the $Q$-factor of the resonator have been observed~\cite{francois2016lasing,franccois2015s,wienhold2015all}. These can lead to an increase in the sensitivity of resonator-based optical sensors~\cite{francois2016lasing,franccois2015fiber}.

Dielectric whispering gallery mode resonators (WGMRs) confine the light via total internal reflection at their rotationally symmetric boundary~\cite{strekalov_nonlinear_2016}.  They are particularly well suited for lasing feedback as the gain material itself acts as the cavity. When these active WGMRs are used together with a suitable pump source, lasing is observed~\cite{garrett1961stimulated}.  A WGMR laser is usually referred to as whispering gallery laser (WGL). WGLs have been demonstrated with resonators made of gain material of different shapes such as toroids~\cite{carmon2005feedback,grudinin2009brillouin}, rings~\cite{chu2011lasing,lacey2007versatile,rong2007low}, microspheres~\cite{min2003compact,ward2011wgm,reynolds2017fluorescent}, asymmetric resonators~\cite{rex_fresnel_2002,schwefel_dramatic_2004,gmachl_high-power_1998,nockel_ray_1997} and disks~\cite{gargas2010whispering,herr_led-pumped_2017}.

\begin{figure*}[hbt!]
  \includegraphics[width=\linewidth]{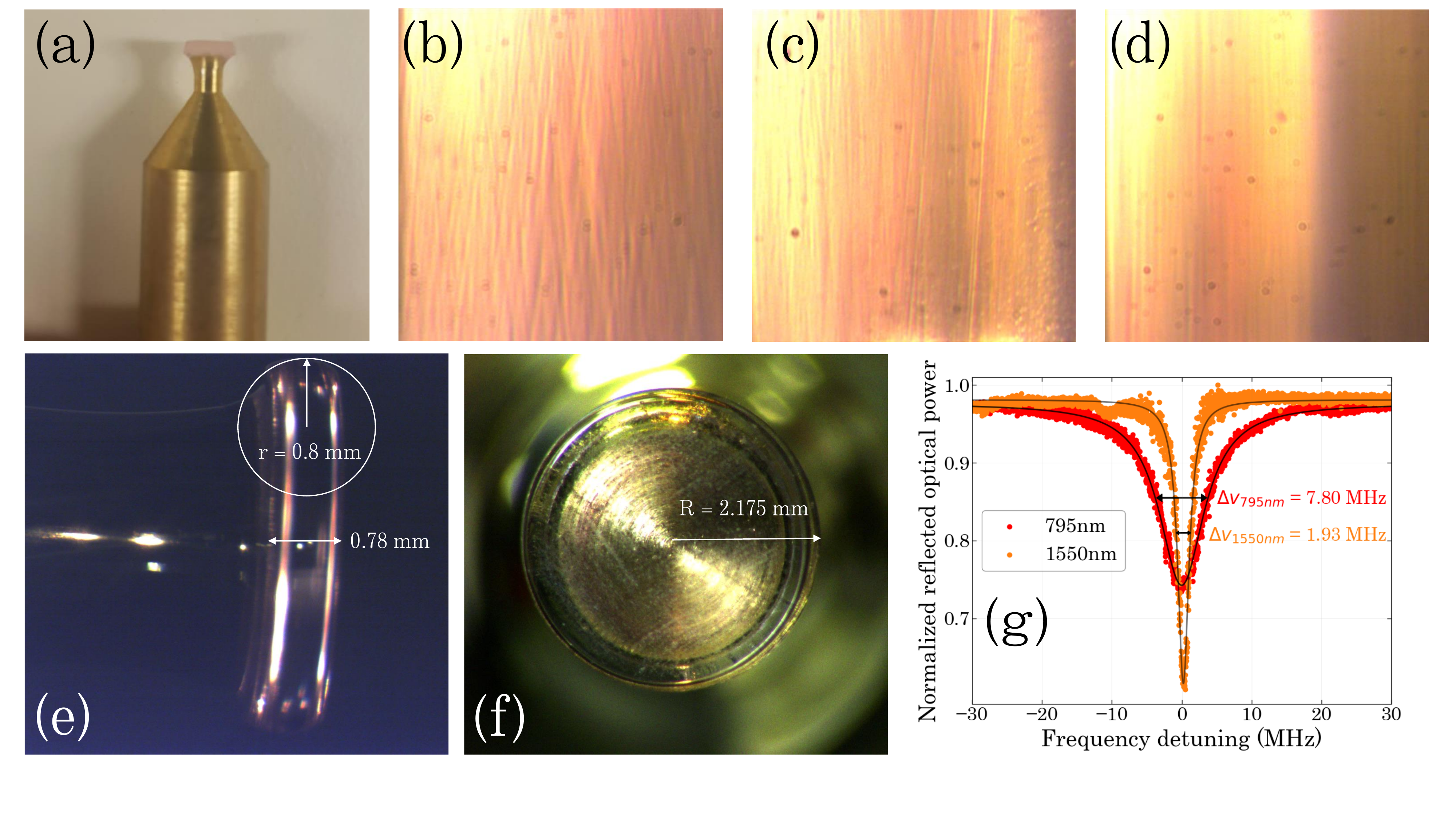}
  \caption{Fabrication process of the Ti:sapphire WGMR. (a) Ti:sapphire crystal is cut into the shape of a disc from a crystal piece using a diamond wheel saw and is then glued on top of a brass rod using mounting wax. (b), (c) and (d) show the surface of the disc after it was polished with diamond slurry of size \SI{30}{\micro\m}, \SI{9}{\micro\m}, and \SI{3}{\micro\m}, respectively. (e) Shows the side view of the finalized resonator indicating its thickness, i.e., \SI{0.78}{\milli\m} and its radius of curvature, $r = \SI{0.80}{\milli\m}$. (f) Shows the top view of the finalized disc, indicating its radius $R = \SI{2.175}{\milli\m}$. (g) The critical linewidths of the modes measured at \SI{795}{\nano\m} shown in red and at \SI{1550}{\nano\m} shown in orange and correspond to critical $Q$-factors of \num{5e7} at and \num{1e8} respectively.}
  \label{fig:Fabrication}
\end{figure*}

As opposed to fabricating a WGMR from an active material to achieve lasing, a passive WGMR can be coated with an active material. A continuous-wave (CW) laser has been demonstrated by coating a polystyrene microsphere by $\text{Tm}^{3+}$-doped energy-looping nanoparticles~\cite{fernandez2018continuous}. Moreover, instead of pumping WGMRs by light, they can be pumped electrically. A semiconductor microdisc laser has been demonstrated by applying InGaAs quantum well-dots as the active area via dry-etching and photo-lithography~\cite{moiseev2018highly}. Finally, there is a separate class of WGLs that use a passive WGMR just as a filter cavity~\cite{liang_miniature_2015,liang_ultralow_2015,sprenger_whispering_2009,collodo_sub-khz_2014,kondratiev_self-injection_2017}.

In this work we demonstrate a disc-shaped Ti:sapphire WGL. Previously, there have been some studies on frequency stability of cryogenic Ti:sapphire WGMRs in the microwave domain~\cite{hartnett_frequency-temperature_1999,hartnett_temperature_1998,boubekeur_frequency_2005}, but, to the best of our knowledge, no study has been performed on lasing in the optical domain. Our WGMR has exhibited the highest recorded critical $Q$-factors at \SI{795}{\nano\m} and \SI{1550}{\nano\m}, \num{5e7} and \num{1e8} respectively, for Ti:sapphire. As a laser pumped at \SI{516.6}{\nano\m} it has an input-power threshold of \SI{14.2}{\milli\W} and a slope efficiency of \SI{34}{\percent} in multi-mode operation, based on the pump power coupled into the resonator at the onset of lasing. This is the lowest observed Ti:sapphire lasing threshold to the best of our knowledge.

\begin{figure*}
  \includegraphics[width=\linewidth]{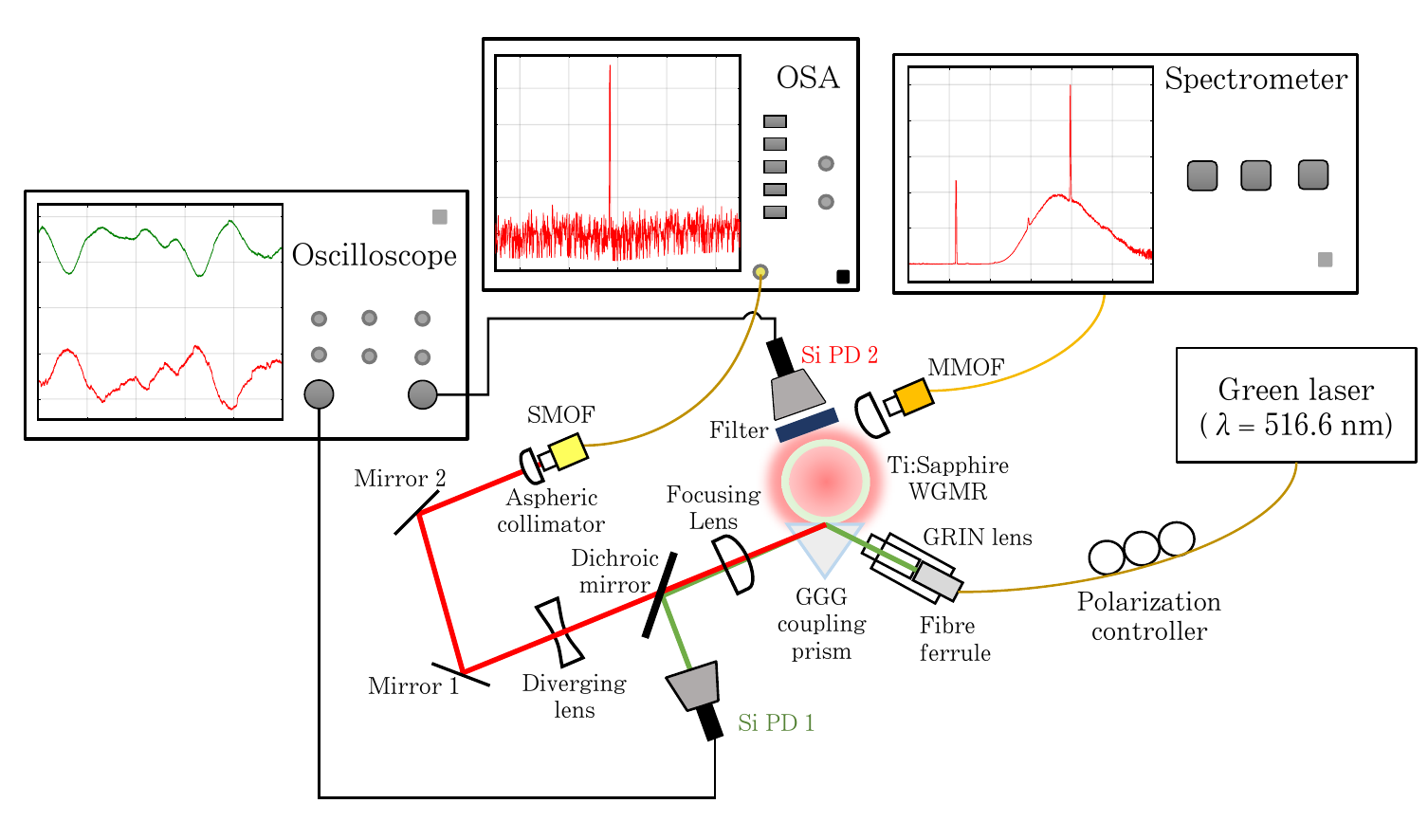}
  \caption{Assembly of apparatus for the lasing experiment. A green ($\lambda = \SI{516.6}{\nano\m}$) laser is coupled into the WGMR using a fibre ferrule, a GRIN lens and a GGG prism. The reflected light from the resonator is split into green and infrared by a dichroic mirror. The green output and the scattered fluorescence/lasing is collected from the resonator using two Si PDs and then monitored using an oscilloscope. A polarization controller and a PBS is used to switch between the TE and TM polarizations. The IR light is coupled into a SMOF using walking the beam method by employing two mirrors and a diverging lens to collimate the light. The scattered signal from the WGMR is also collected using a MMOF and monitored using a spectrometer.}
  \label{fig:Lasingsetup}
\end{figure*}

\section{Experiments}

We fabricated a single crystalline Ti:sapphire WGMR as shown in Figure~\ref{fig:Fabrication}(a-f) (see  Methods section~\ref{sec:methodsFabrication}). Light was coupled into the resonator evanescently through prism coupling~\cite{strekalov_nonlinear_2016} (see Figure~\ref{fig:Lasingsetup} and Methods section~\ref{sec:LasingSetup}). First we determined the $Q$-factor around \SI{795}{\nano\m} and \SI{1550}{\nano\m} by scanning two different tunable diode lasers over a resonance while changing the coupling distance of the prism (see Supplement), measuring a critical $Q$-factor of \num{5e7} and \num{1e8}, respectively (see Figure~\ref{fig:Fabrication}(g)). The lower $Q$-factor at \SI{795}{\nano\m} is expected due to the re-absorption of titanium in the WGMR, and not due to the surface quality. Both $Q$-factors are slightly lower than reported in the study performed by~\cite{ilchenko2014generation} on undoped sapphire. In order to observe lasing, we coupled a tunable green laser ($\lambda = \SI{516.6}{\nano\m}$) evanescently into the WGMR. The resonator fluorescence  was detected by imaging a segment of the resonator's rim into a spectrometer, while the near-infrared (IR) output was observed through the coupling prism using an optical spectrum analyzer (OSA). The rim fluorescence was also detected by a photodiode. The $Q$-factor in the green is of the order of \num{5e5} (see Supplement), as expected for the strong absorption of the titanium dopant.

\begin{figure*}
  \includegraphics[width=0.3\linewidth]{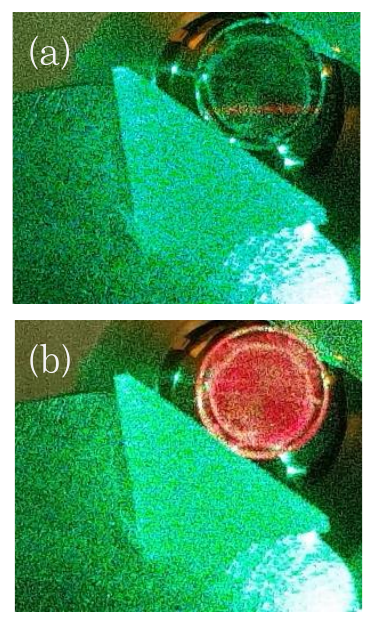}
  \includegraphics[width=0.657\linewidth]{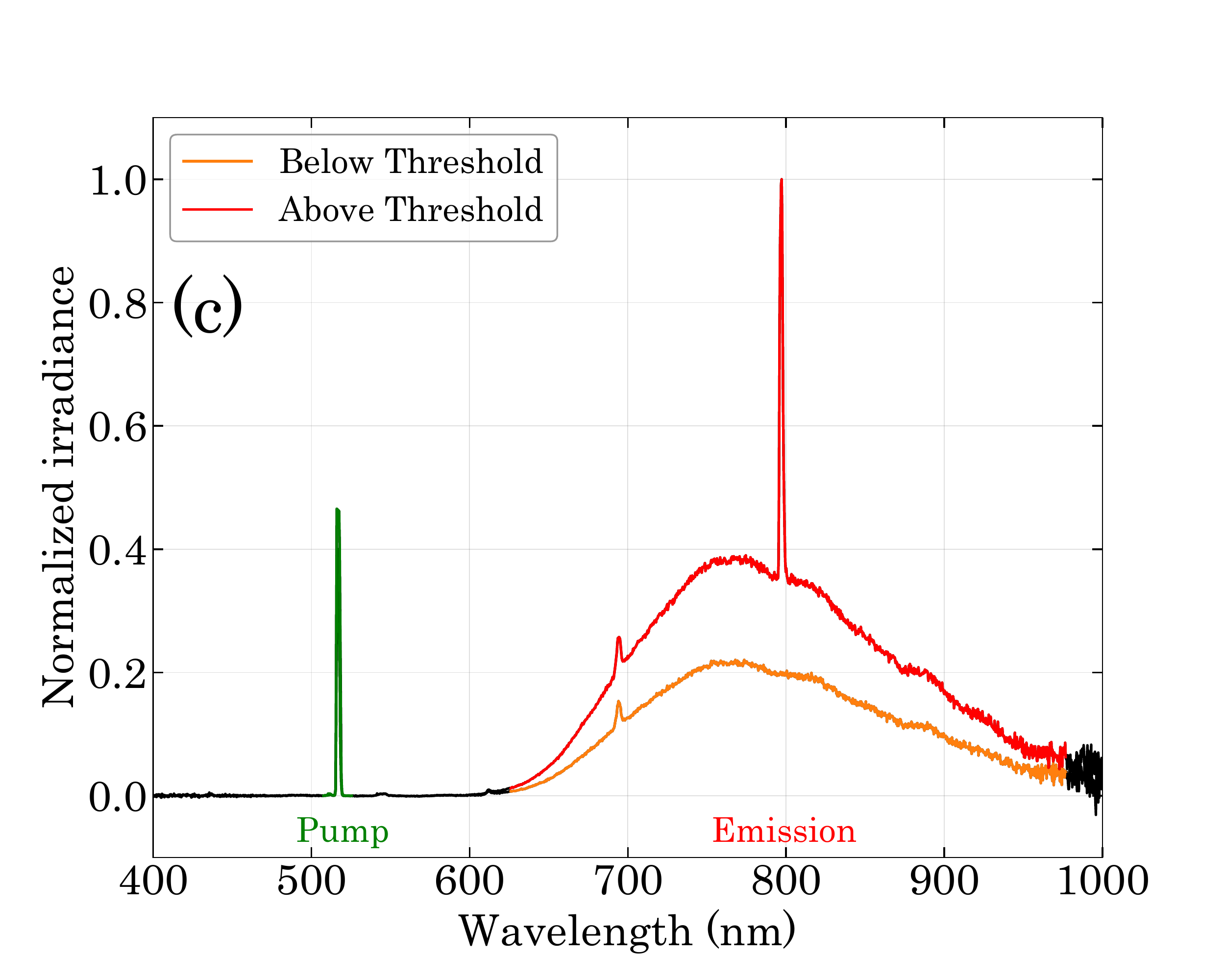}
  \caption{Fluorescence and lasing in the Ti:sapphire WGL. (a) Shows the Ti:sapphire resonator and the coupling prism. (b) Shows the resonator when the green pump is coupled to the WGMR. The reddish glow in the resonator is due to the gain excited in the resonator. (c) Comparison of the resonator fluorescence spectra collected using the spectrometer. The orange trace shows the emission when the pump power is below threshold and the red trace shows the emission when the threshold is exceeded.}
  \label{fig:Fluorescence}
\end{figure*}

Lasing in the WGMR can be either multi-mode or single-mode. Different lasing modes can be excited by changing parameters in the experimental setup such as the pump coupling angle, its polarization, coupling strength controlled by the distance between the prism and the resonator, and the input power of the pump laser. By varying these parameters we can excite different pump WGMs in the resonator which gives rise to different lasing modes. 
Once the pump laser is coupled into the resonator at a high enough efficiency, fluorescence is excited in the WGMR. The fluorescence inside the resonator can be observed with the naked eye and photographed, as shown in Figure~\ref{fig:Fluorescence}. Figure~\ref{fig:Fluorescence}(a) shows the WGMR when there is a significant distance between the prism and the resonator, i.e., the green light is not coupled into the WGMR. Figure~\ref{fig:Fluorescence}(b) shows the resonator when the green light is coupled to the resonator. The fluorescence inside the resonator can be observed as a bright reddish glow. The first observation of lasing is shown in Figure~\ref{fig:Fluorescence}(c), which shows the pump and the emission spectra of the imaged resonator; the orange line shows the emission data collected by the spectrometer when the lasing threshold is not exceeded, whereas the red line shows the same when the lasing threshold is exceeded. A lasing peak is clearly seen near \SI{797}{\nano\m}. The small peak observed at \SI{695}{\nano\m} is due to traces of chromium (ruby) doping~\cite{mahnke2001aluminum}.

\begin{figure*}
  \centering
  \includegraphics[width=0.32\linewidth]{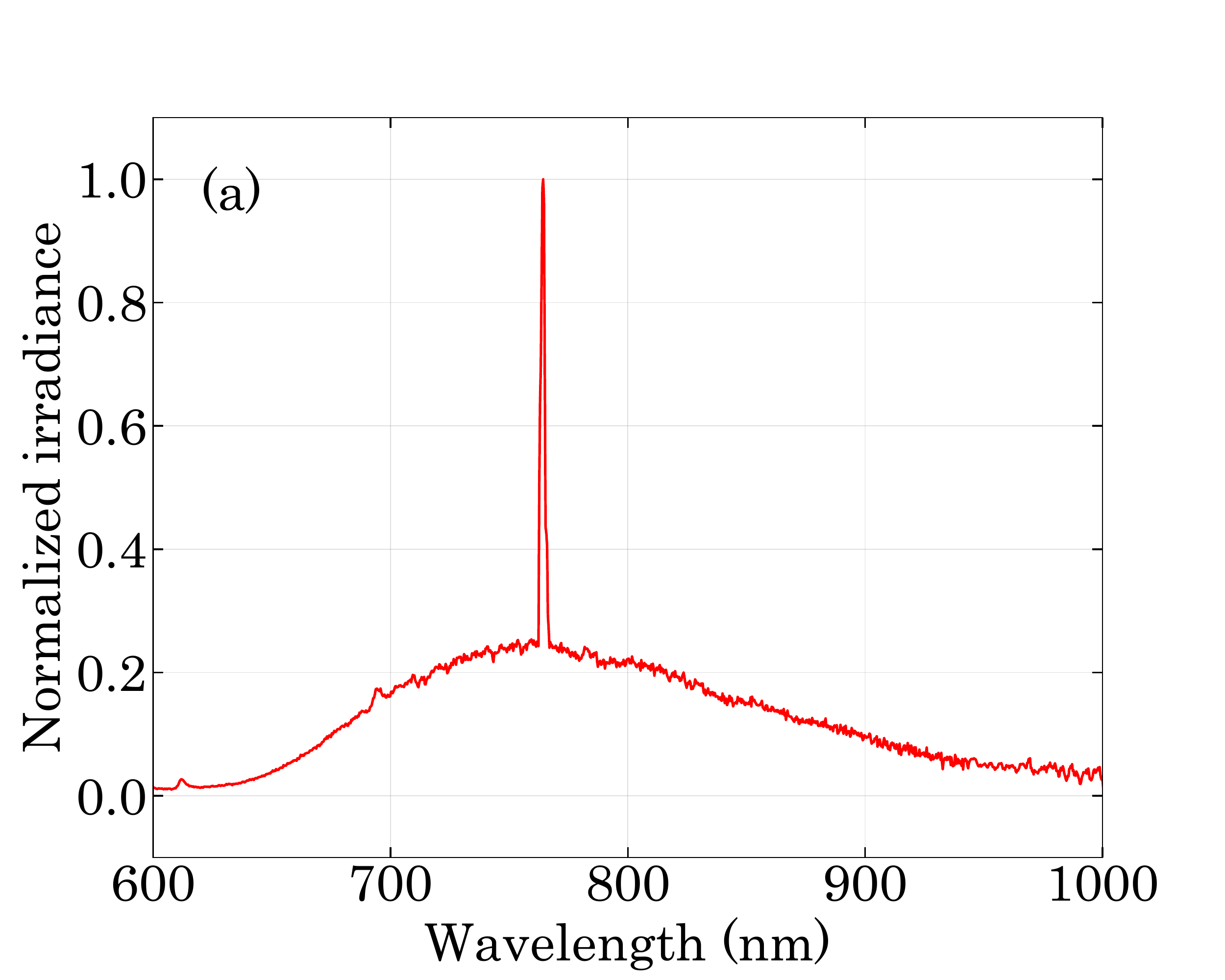}
  \includegraphics[width=0.32\linewidth]{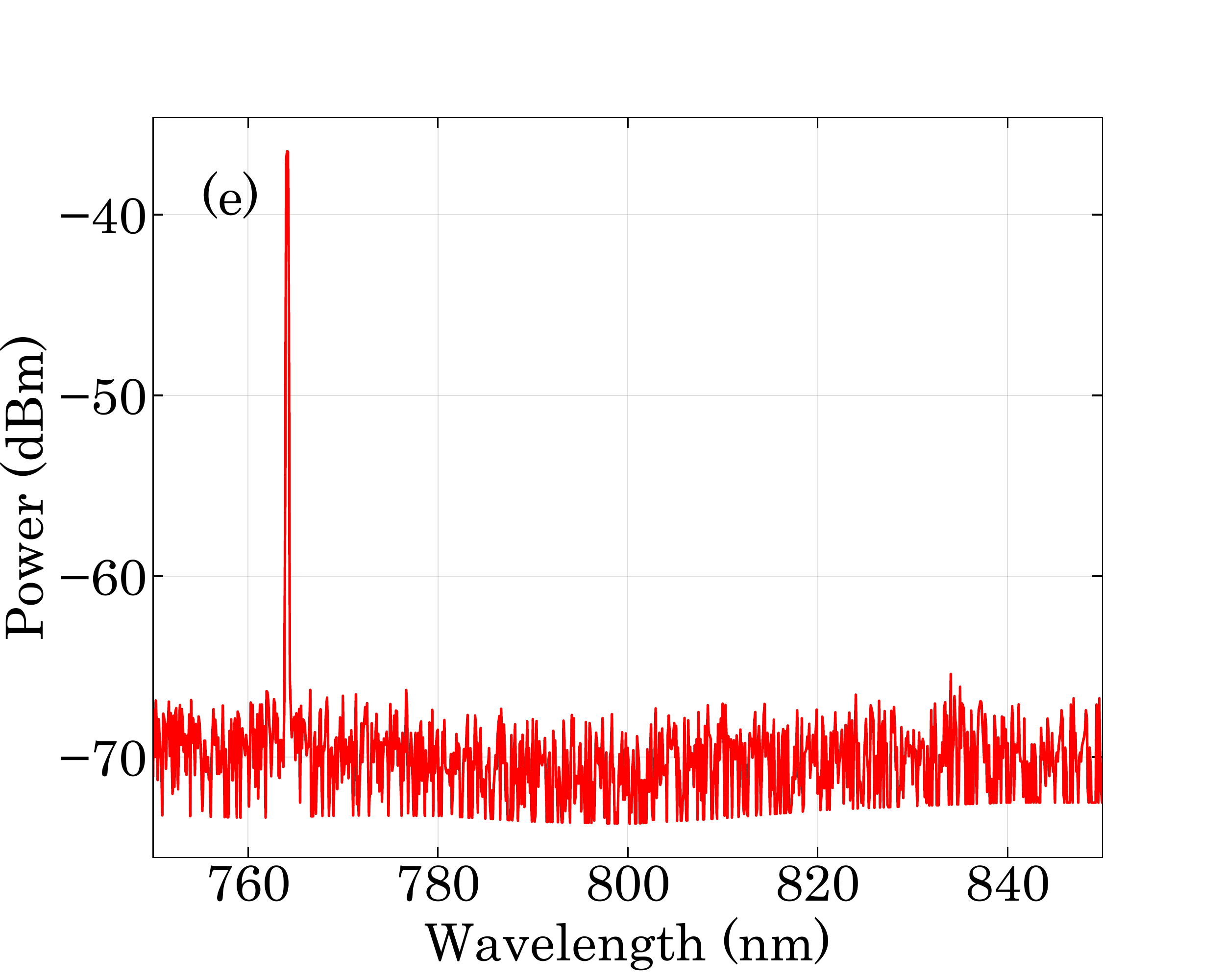}
   \includegraphics[width=0.32\linewidth]{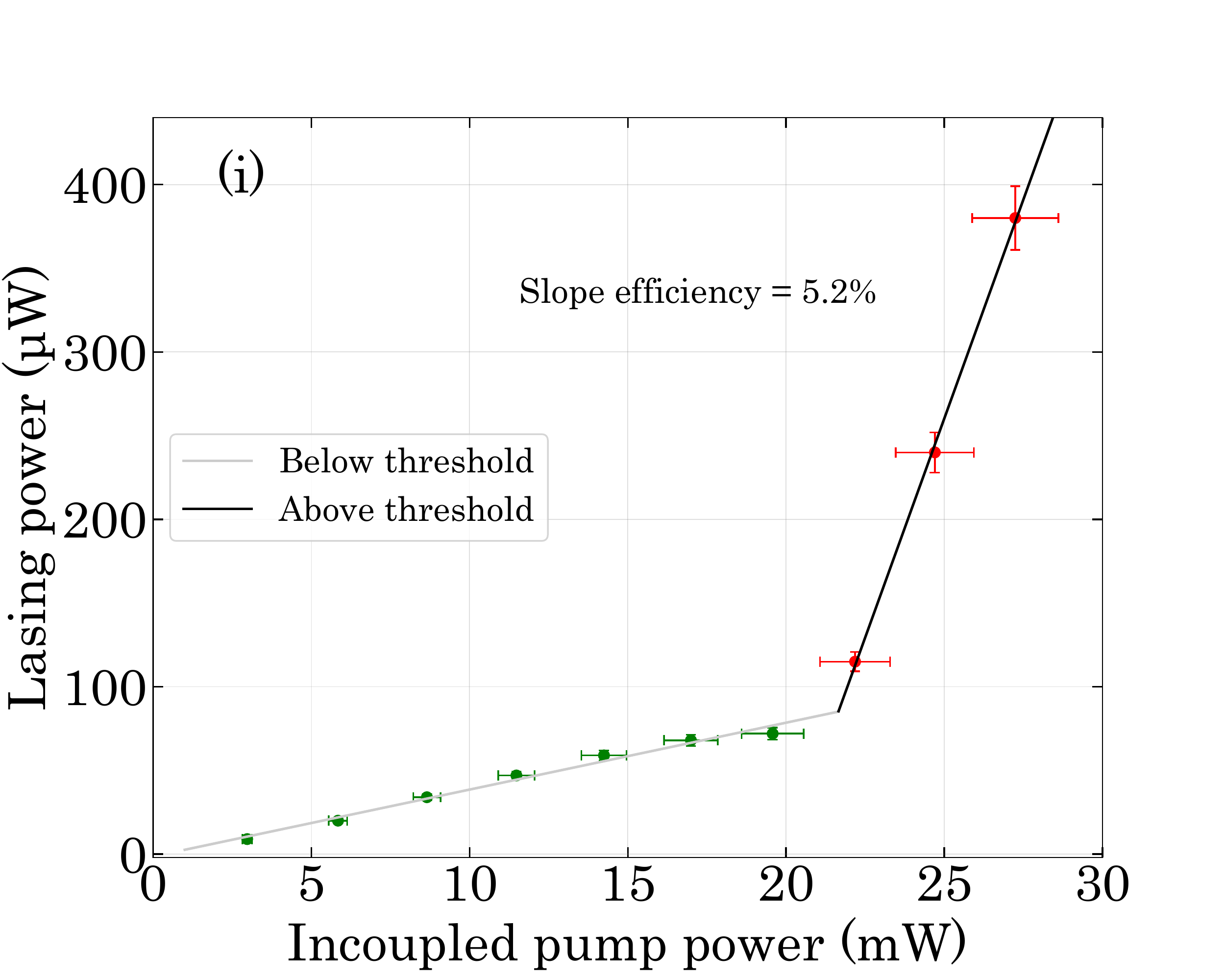}\\
  \includegraphics[width=0.32\linewidth]{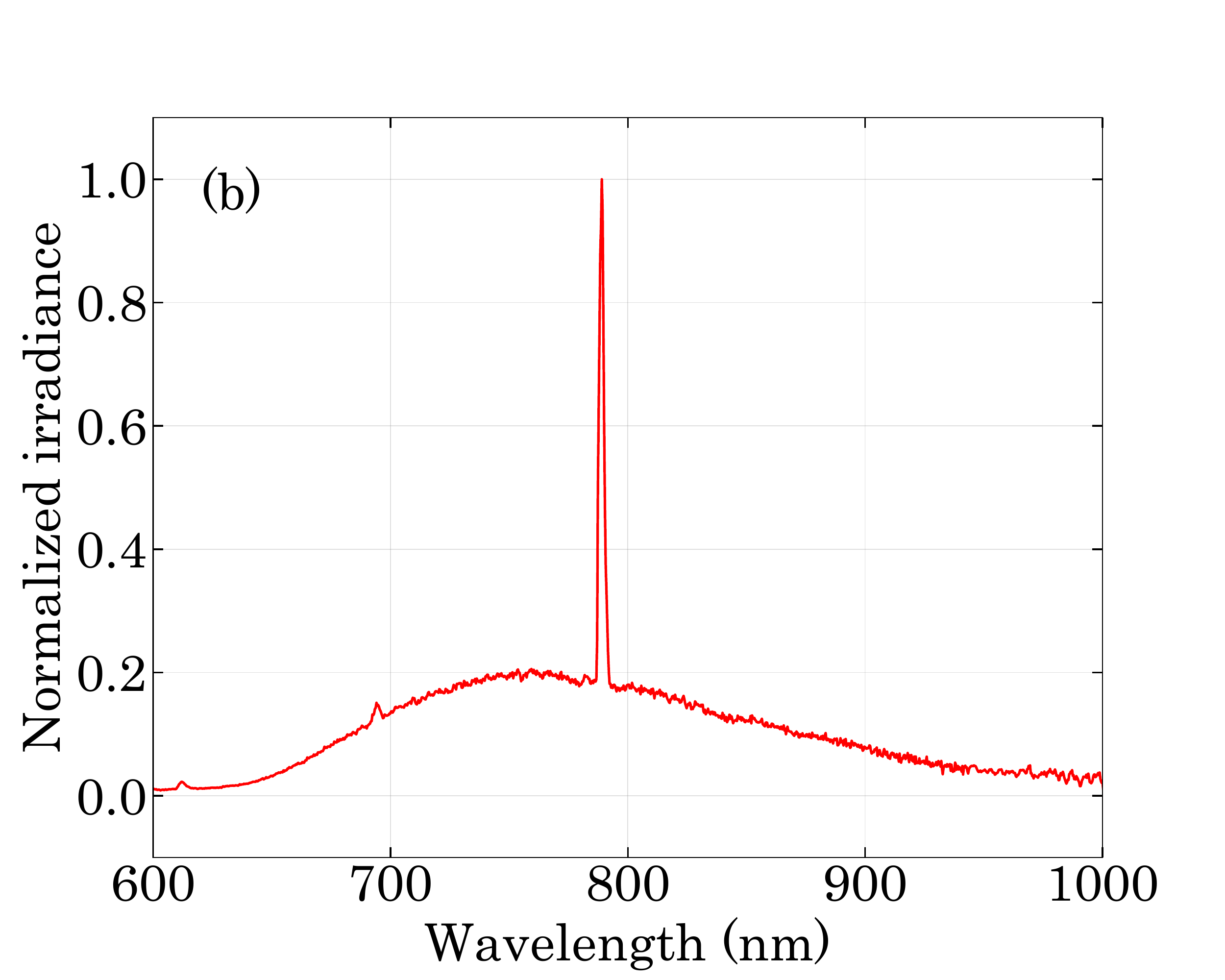}
  \includegraphics[width=0.32\linewidth]{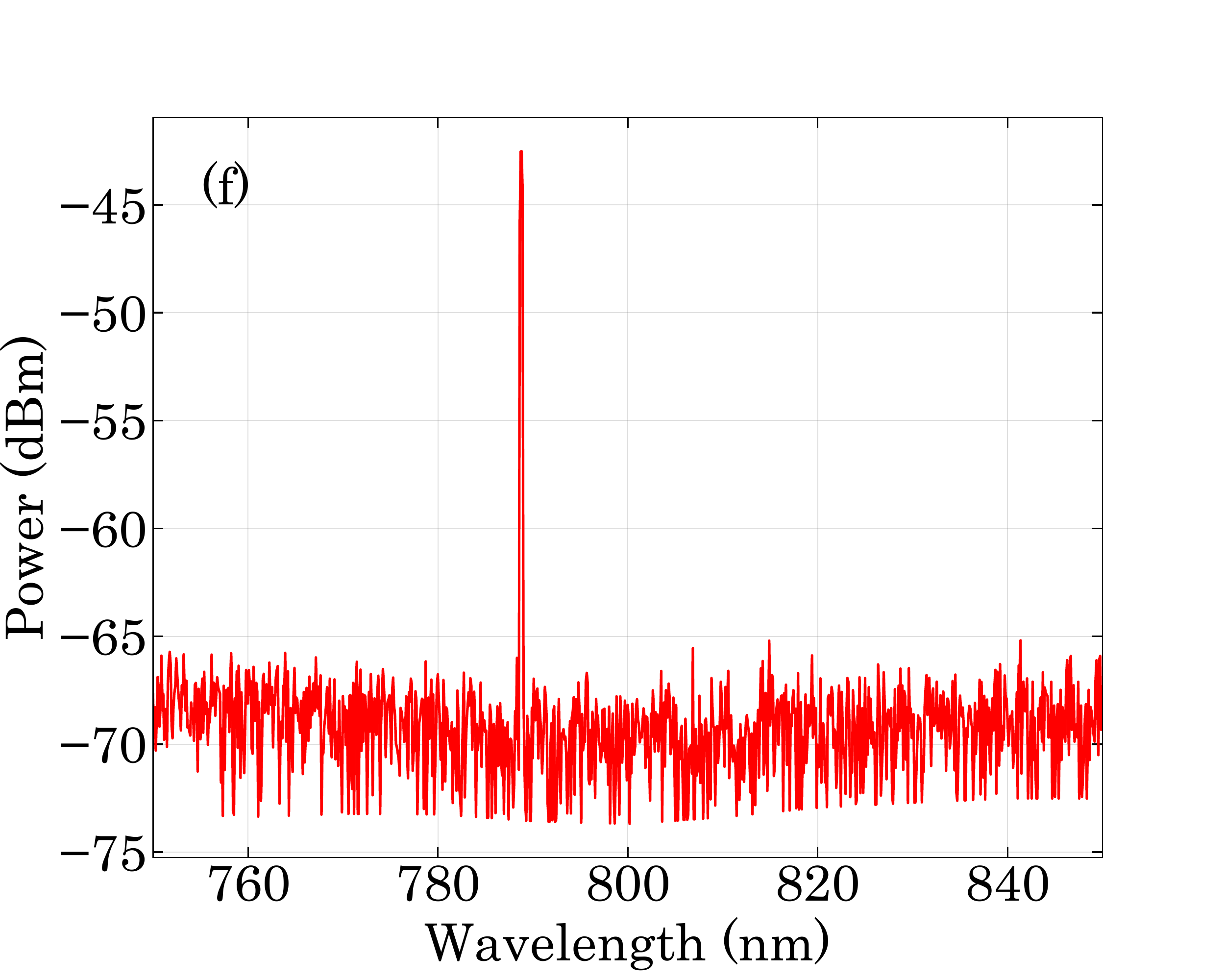}
    \includegraphics[width=0.32\linewidth]{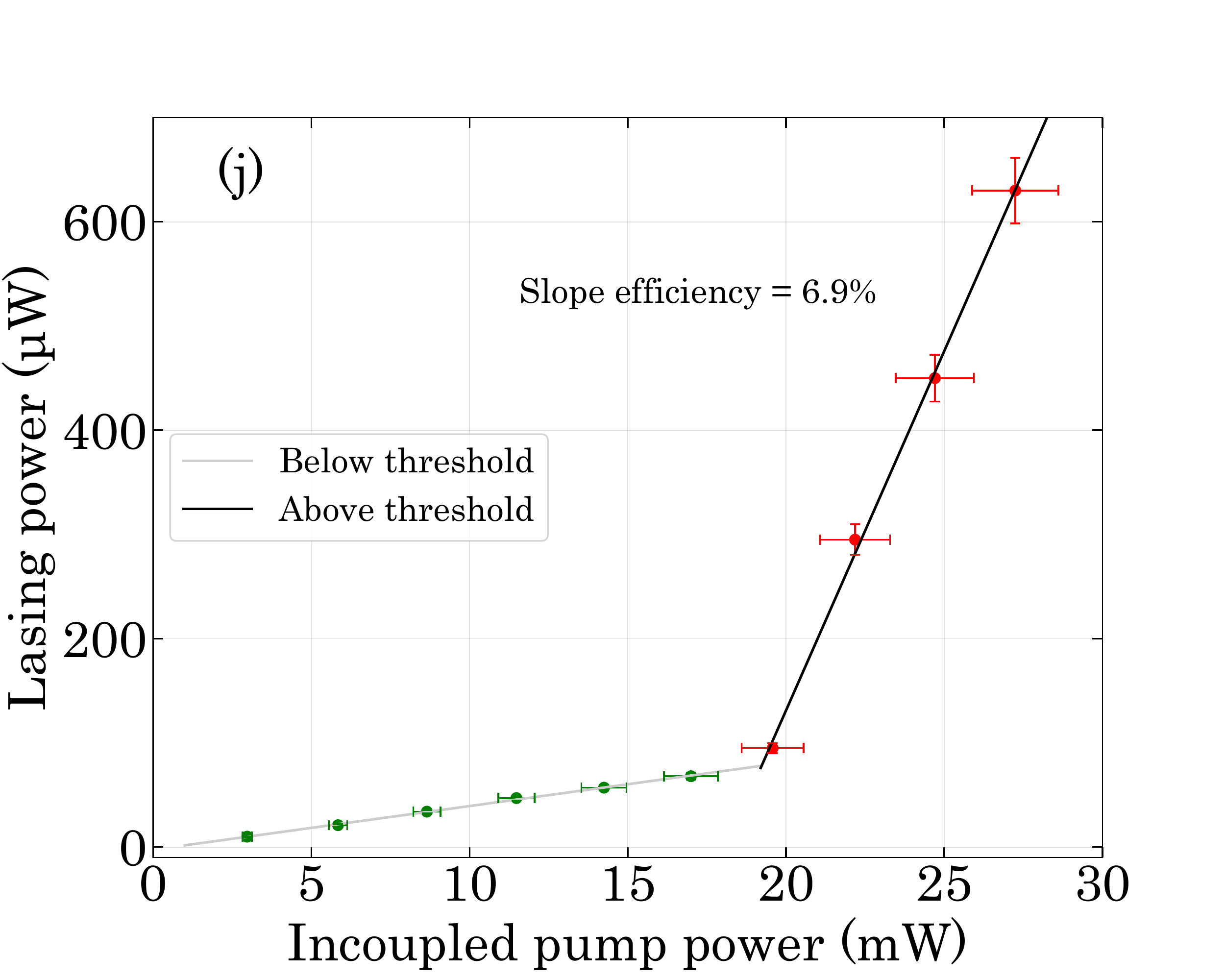}\\
    \includegraphics[width=0.32\linewidth]{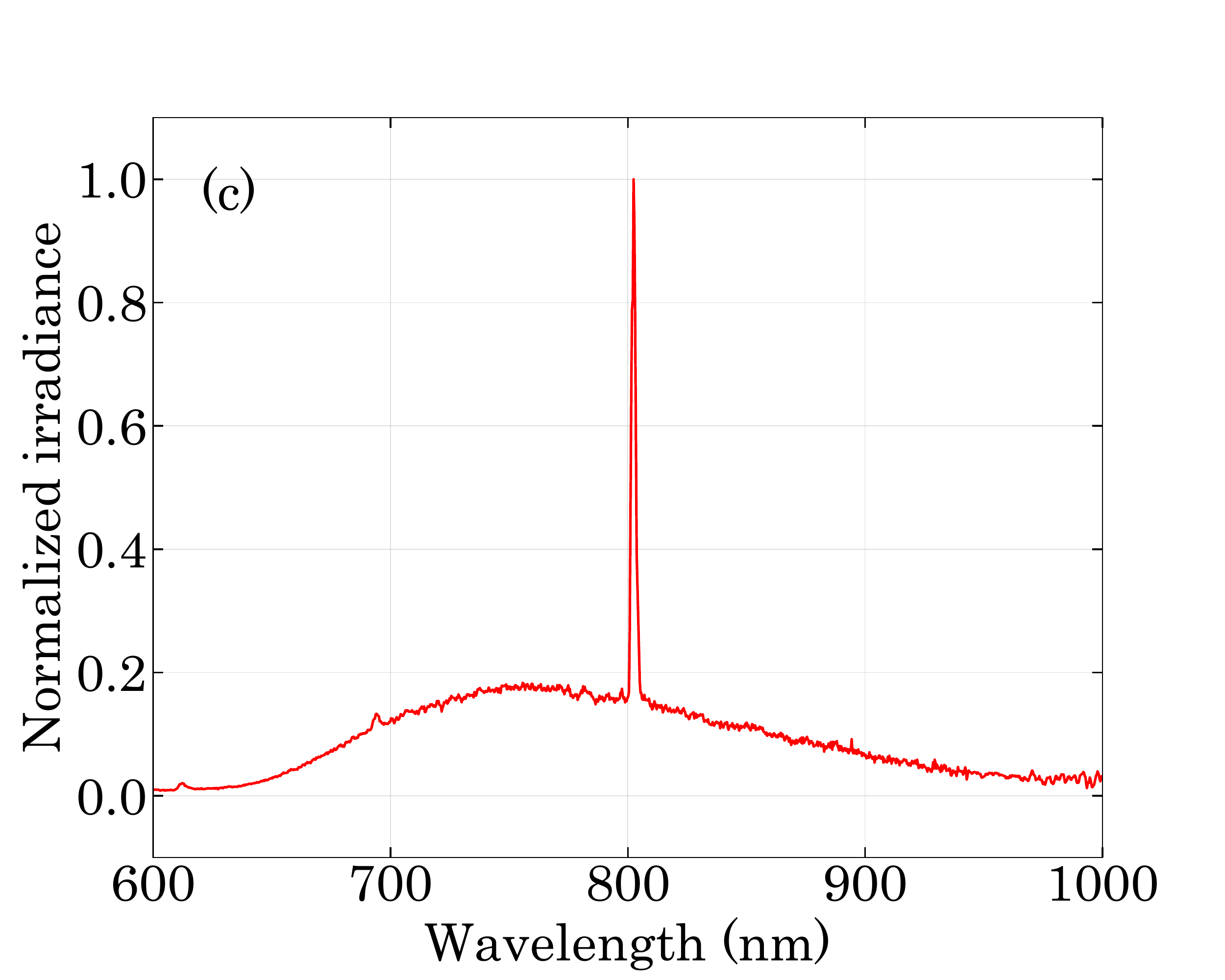}
  \includegraphics[width=0.32\linewidth]{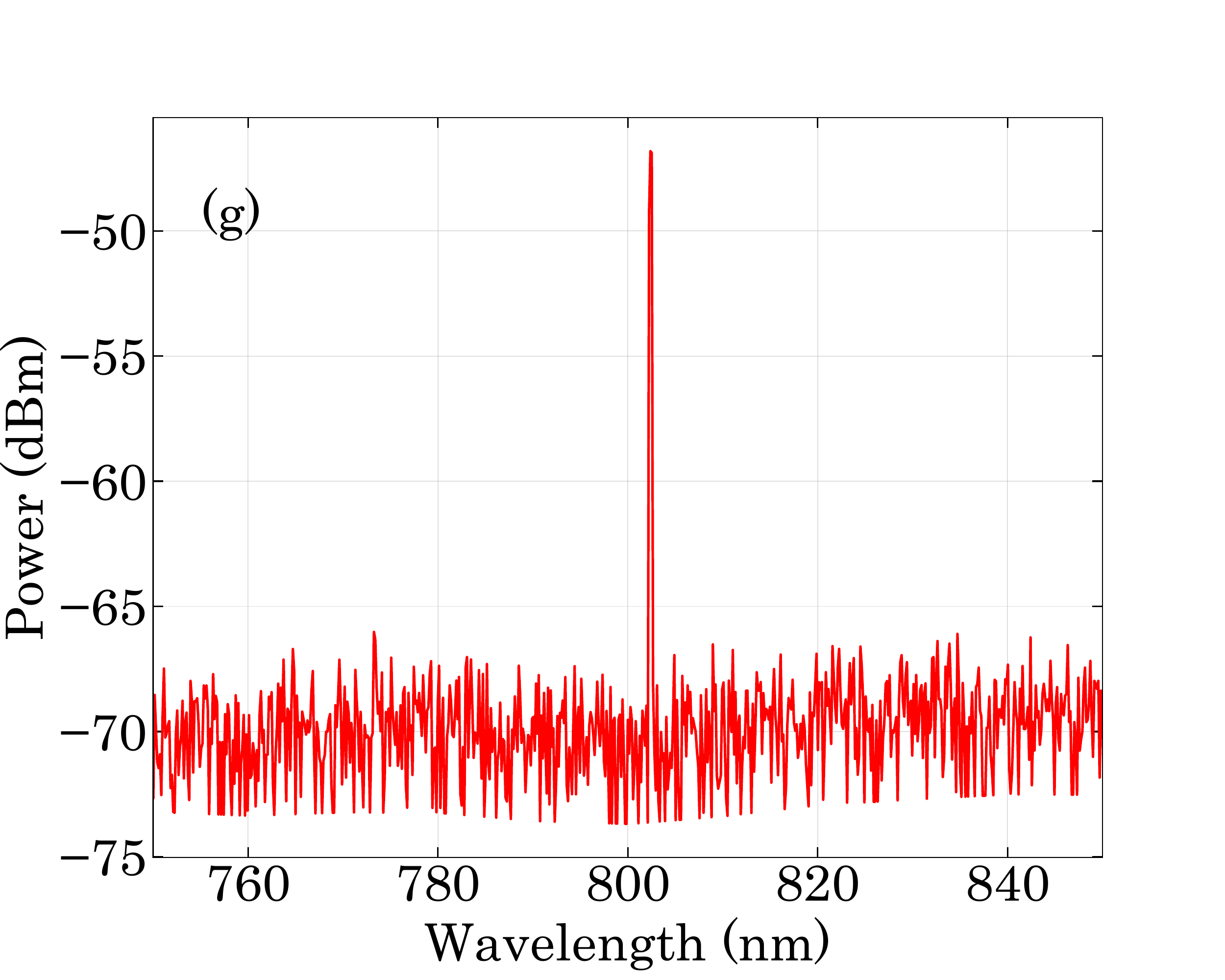}
  \includegraphics[width=0.32\linewidth]{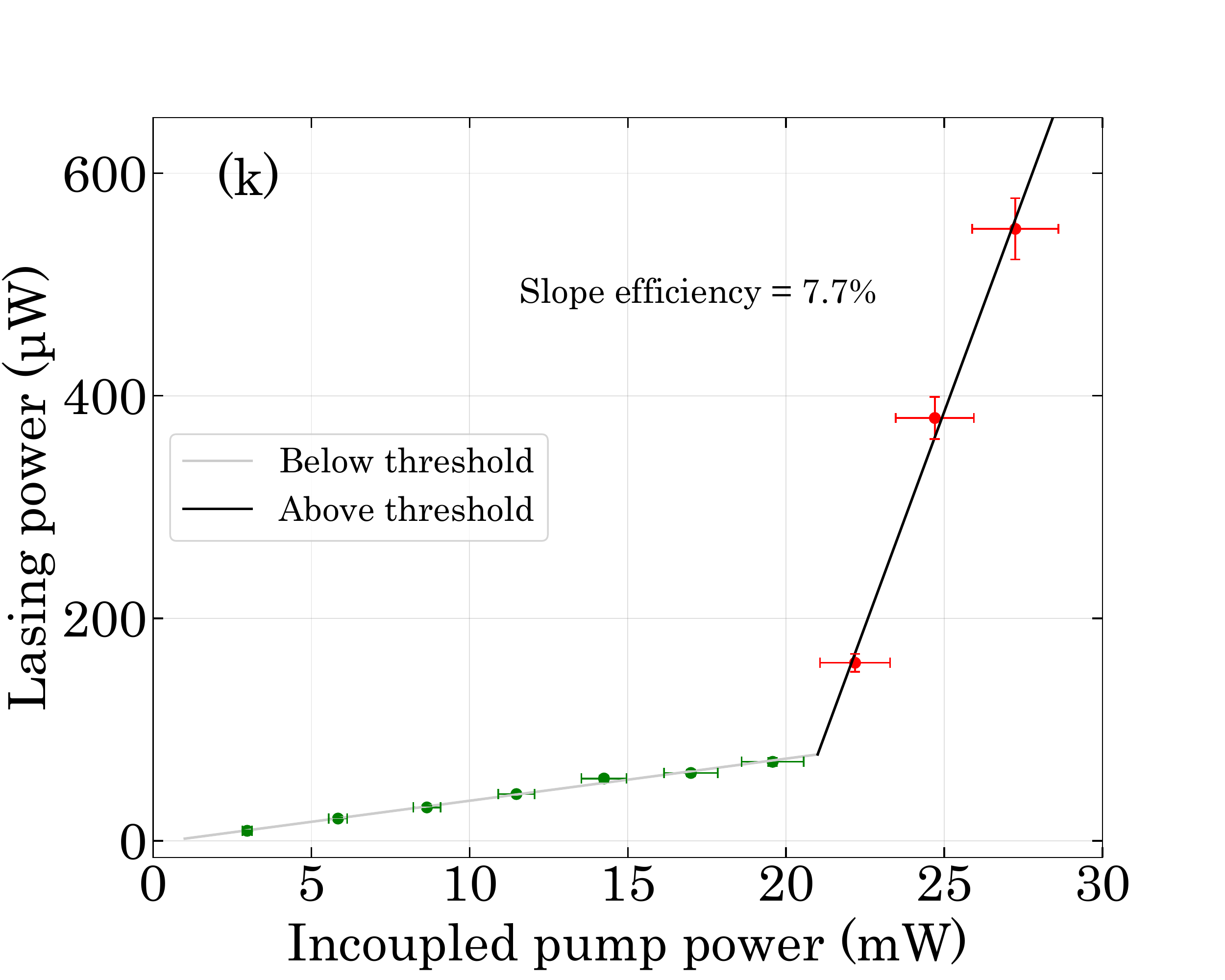}\\
    \includegraphics[width=0.32\linewidth]{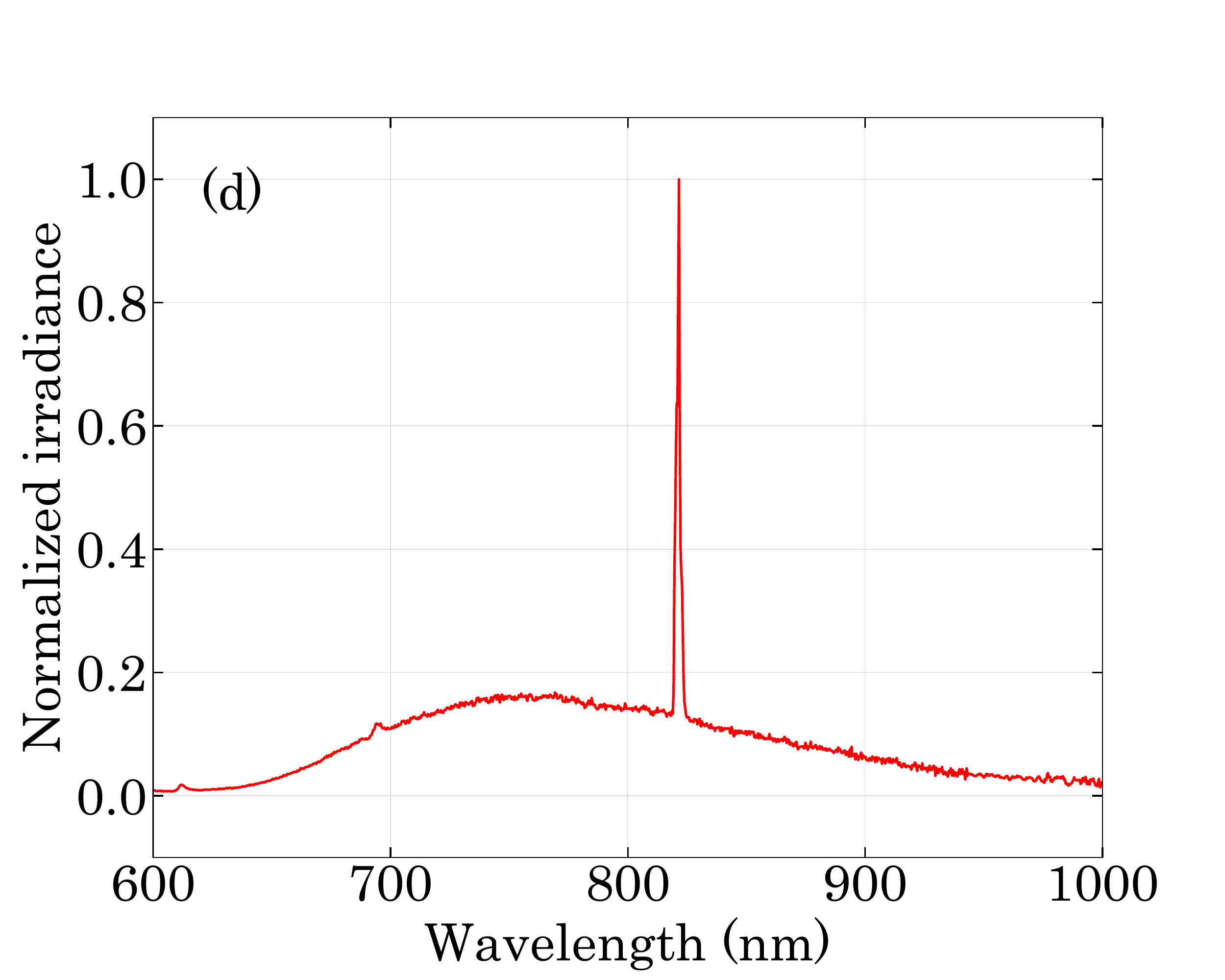}
  \includegraphics[width=0.32\linewidth]{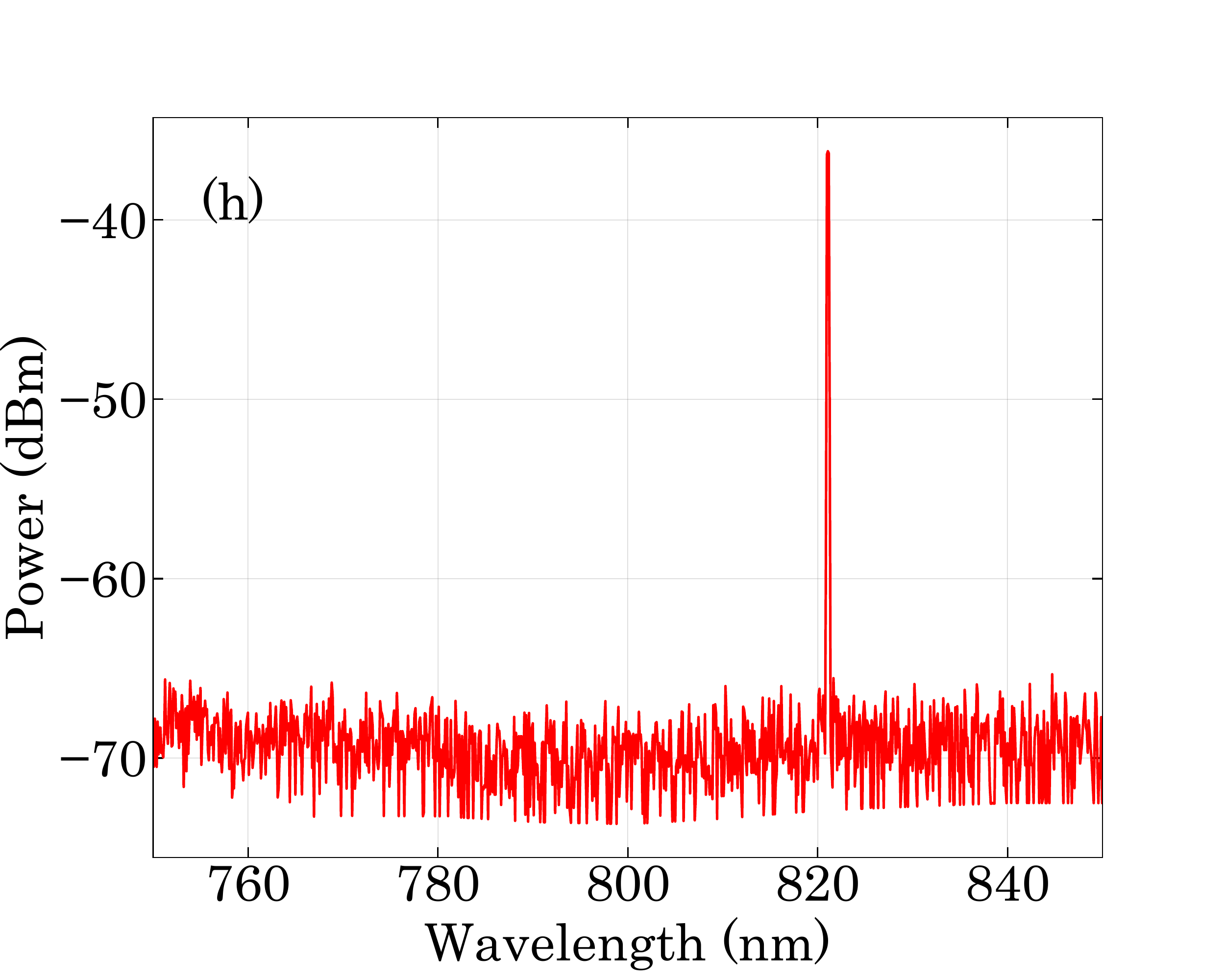}
  \includegraphics[width=0.32\linewidth]{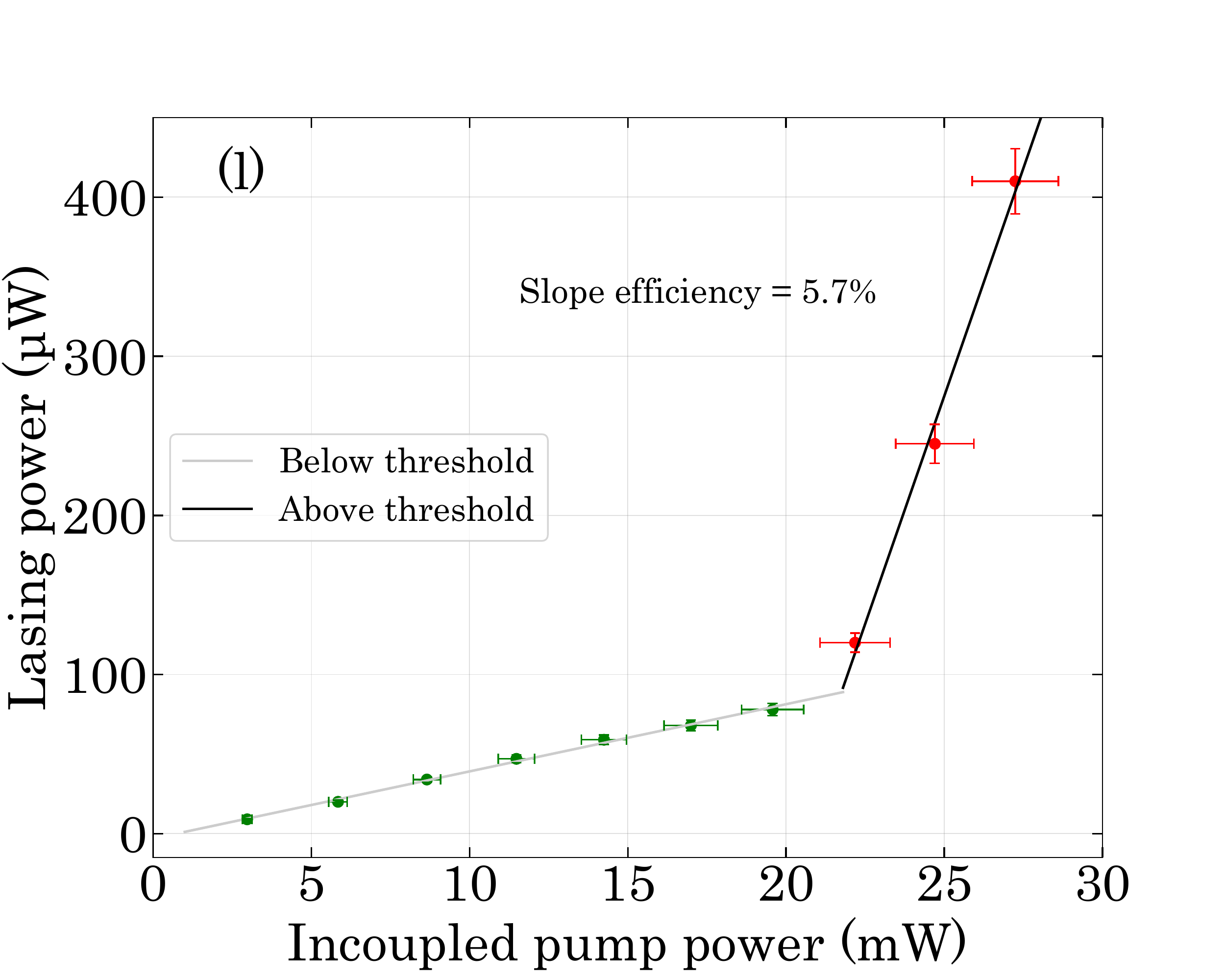}\\
  \caption{Various single-mode lasing peaks recorded.
  (a), (b), (c) and (d) show the single lasing modes in the Ti:sapphire WGL at  \SI{764.1}{\nano\m}, \SI{788.8}{\nano\m}, \SI{802.4}{\nano\m} and \SI{821.1}{\nano\m} recorded via the spectrometer. (e), (f), (g) and (h) show the OSA data in the same order for the aforementioned modes, whereas (i), (j), (k) and (l) show show the output lasing powers versus the incoupled green input pump for these lasing modes.}
  \label{fig:SMLasing}
\end{figure*}

We reach the single-mode regime by adjusting the pump mode so that it favours one lasing mode. This process is know to be highly selective \cite{tureci_theory_2007,ge_quantitative_2008,aung_threshold_2015}, depending on the actual spatial profile of the pump mode. Various single-mode lasing peaks were observed during the experiment. Figures~\ref{fig:SMLasing}(a), (b), (c) and (d) show the spectrometer data for different lasing modes, whereas Figures~\ref{fig:SMLasing}(e), (f), (g) and (h) show the OSA data for the same  modes. Figures~\ref{fig:SMLasing}(i), (j), (k) and (l) show the lasing threshold for these lasing modes observed at the central wavelengths of \SI{764.1}{\nano\m}, \SI{788.8}{\nano\m}, \SI{802.4}{\nano\m}, and \SI{821.1}{\nano\m}, respectively, by varying the green pump power from zero to the full power achievable. The theoretical model predicts two different linear slopes below and above the threshold~\cite{koechner2006solid}. We obtain the above-threshold slope efficiencies for these different lasing modes of 5.2\%, 6.9\%, 7.7\% and 5.7\%, respectively. The exact lasing thresholds are also different for the shown modes. The slope efficiencies and thresholds are expected to be different for different modes due to the intricate interplay between gain profile, pump mode, pump power, coupling efficiency, and gain competition. As the OSA resolution is limited to \SI{28}{\giga\hertz} at the lasing wavelengths, we have confirmed single mode lasing by observing no beat signal of the lasing mode on a fast photo detector (bandwidth \SI{20}{\giga\hertz}), which is larger than one free spectral range.

\begin{figure*}
  \centering
  \includegraphics[width=0.49\linewidth]{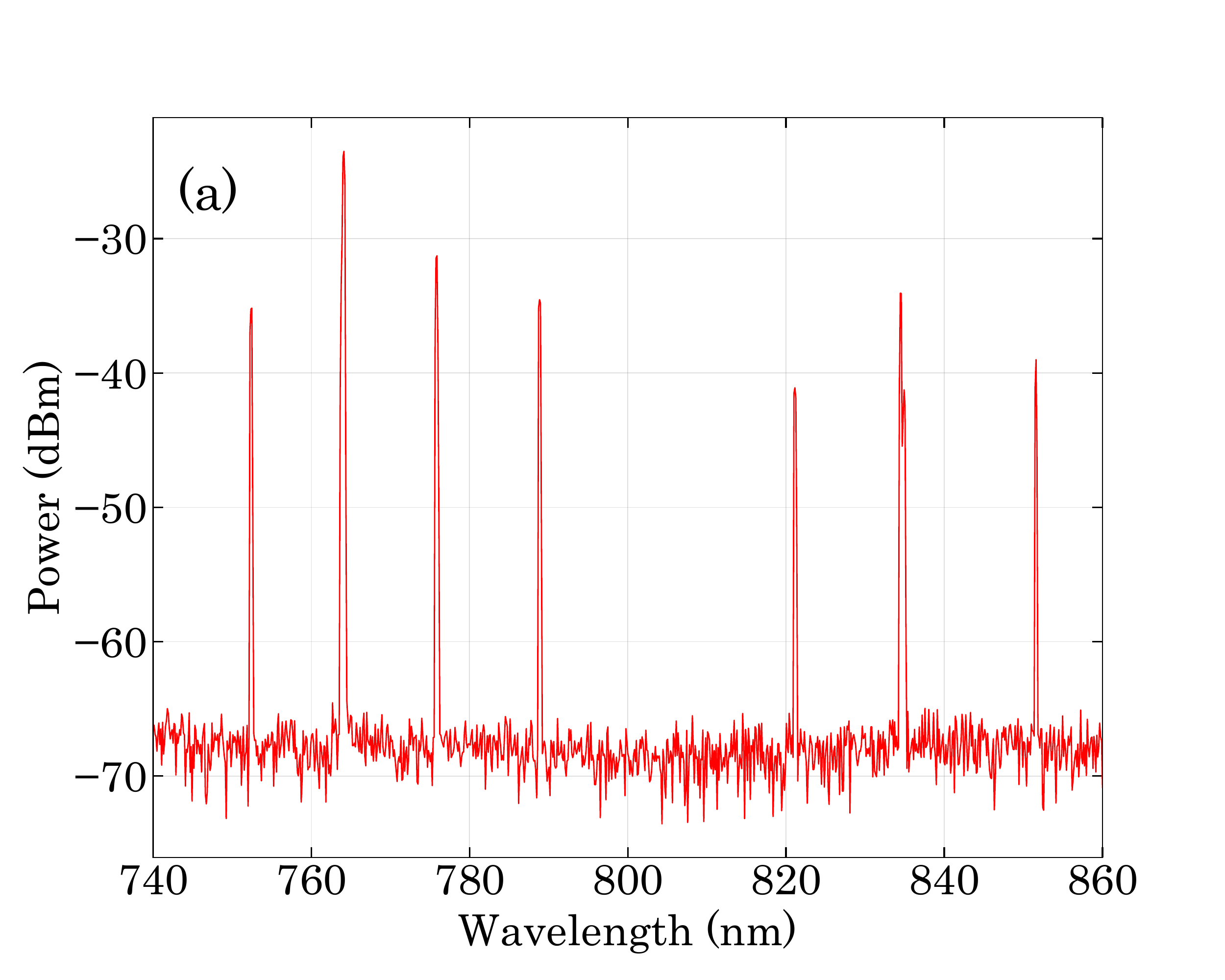}
  \includegraphics[width=0.49\linewidth]{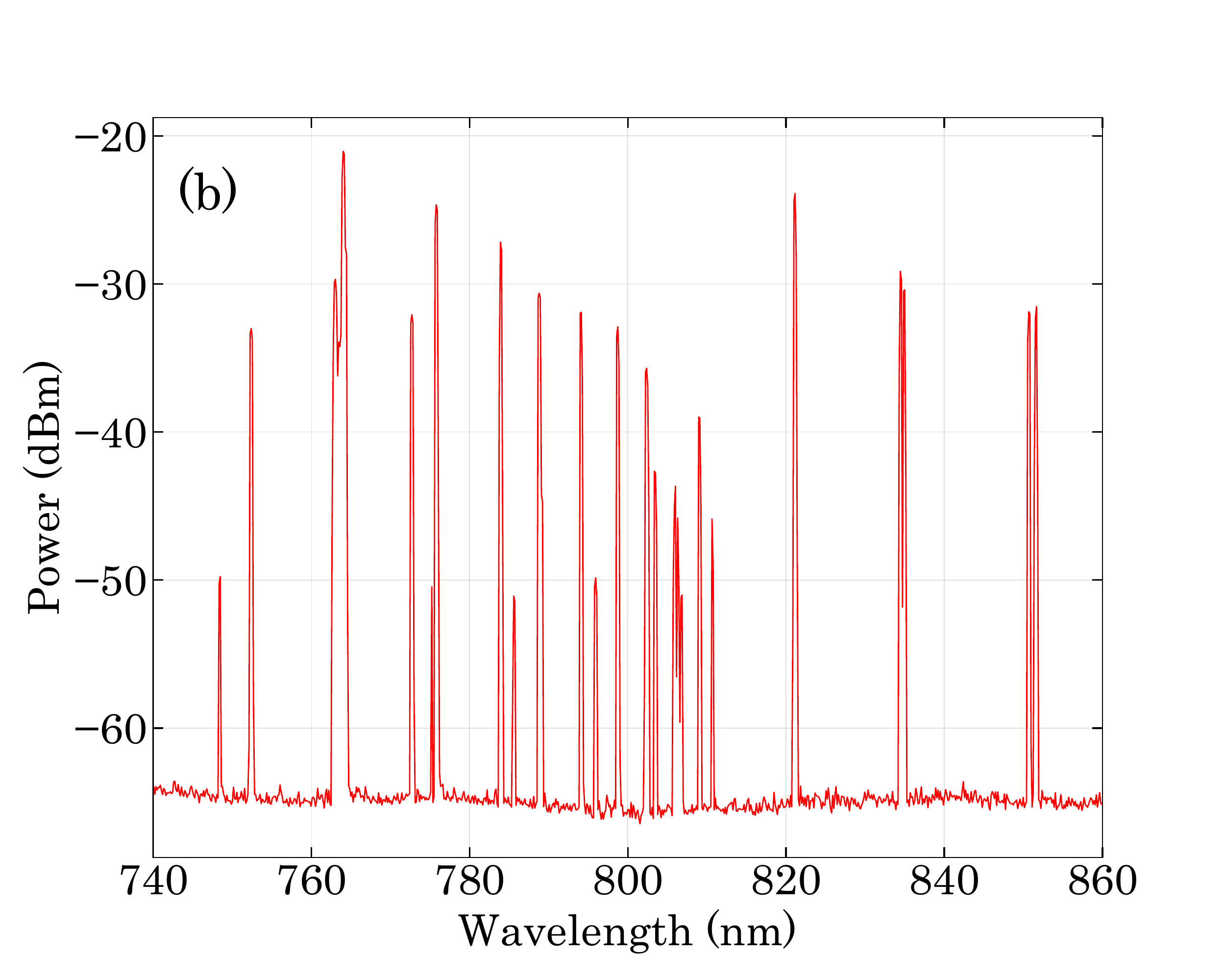}\\
  \includegraphics[width=0.49\linewidth]{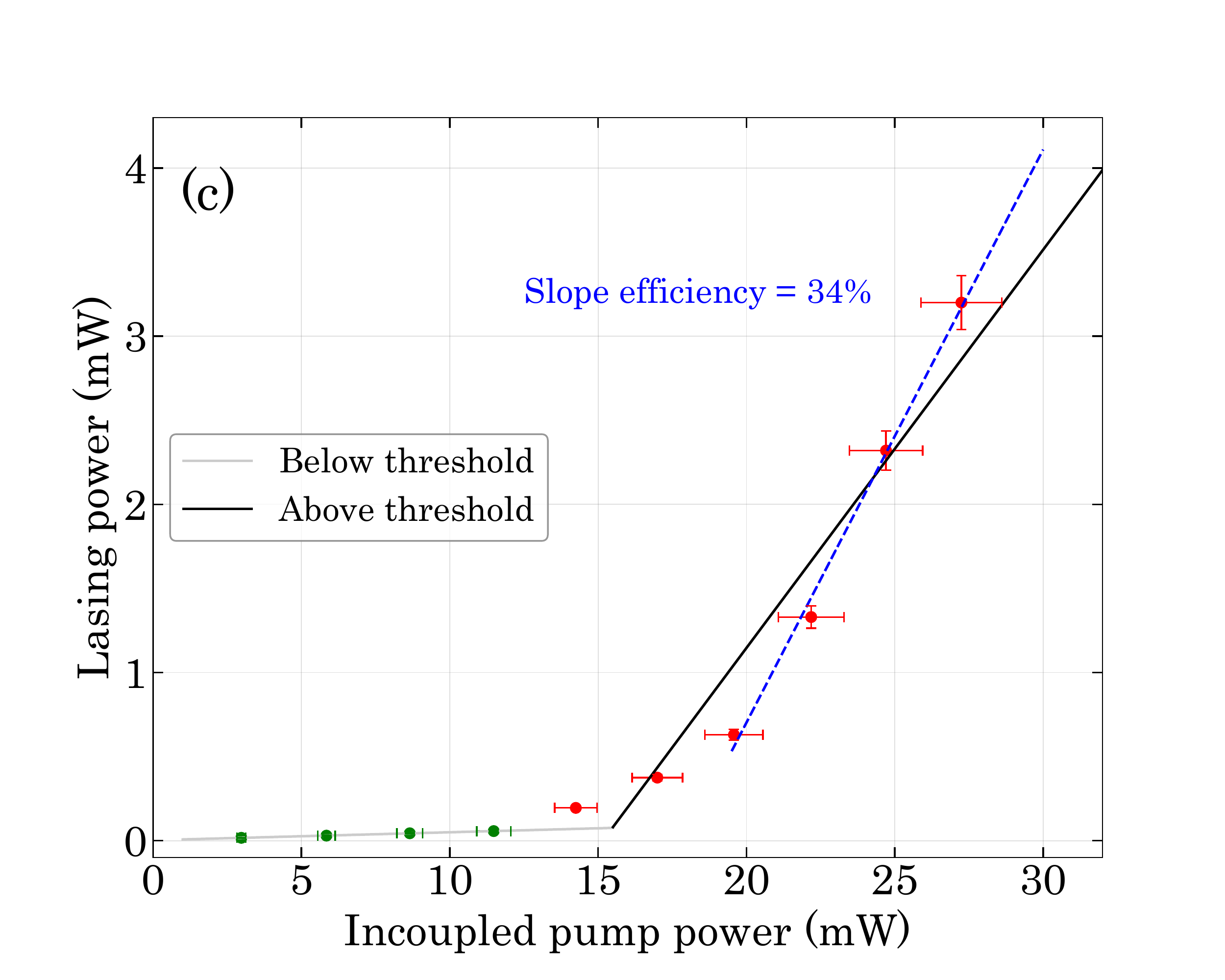}
  
  \caption{Lasing in Ti:sapphire WGL.  (a) Multi-mode lasing observed in a single instance spanning a range of almost \SI{100}{\nano\m}. A minimum of seven different lasing peaks are observable in the plot at different wavelengths. (b) Multiple lasing peaks observed at various wavelengths. The maxhold function on the OSA was used to record these lasing peaks while varying the distance between the prism and the resonator, the input polarization, the input power, and the pump laser frequency. (c) Output lasing power versus input pump power for the multi-mode lasing in (a). The observed lasing threshold for multi-mode lasing is at \SI{14.2}{\milli\W}, with a maximum slope efficiency of 34\%. The error bars are a result of the $\pm5\%$ measurement accuracy of the Thorlabs power meter.}
  \label{fig:MMLasing}
\end{figure*}

Multi-mode lasing in our WGL is typically triggered when different lasing modes do not compete with each other. Figure~\ref{fig:MMLasing}(a) shows the multi-mode lasing peaks observed in a single OSA sweep. At least seven different lasing peaks are observable in the plot at wavelengths of \SI{752.4}{\nano\m}, \SI{764.1}{\nano\m}, \SI{775.9}{\nano\m}, \SI{788.8}{\nano\m}, \SI{821.1}{\nano\m}, \SI{834.6}{\nano\m}, and \SI{851.6}{\nano\m}, spanning a range of almost \SI{100}{\nano\m}. Figure~\ref{fig:MMLasing}(b) shows all of the multi-mode lasing peaks recorded using the max hold function of the OSA. These peaks were recorded while parameters such as the pump power, polarization, and coupling conditions were changed. The shortest wavelength lasing peak recorded in the figure is at \SI{748.5}{\nano\m}, whereas the longest wavelength lasing peak is at \SI{851.6}{\nano\m}. This confirms that the Ti:sapphire WGL has a range spanning more than \SI{100}{\nano\m}.

The cumulative lasing power versus the input pump power for multi-mode lasing is plotted in Figure~\ref{fig:MMLasing}(c). For this the power of the green pump laser is varied from zero to the full power achievable. At the optimal coupling conditions, i.e., coupling angles and coupling rates, the onset of lasing occurs at  \SI{14.2}{\milli\W}. We attribute this to the excitation of the first lasing WGM which has the lowest threshold. As the pump power increases, more WGMs with higher thresholds begin to lase, which leads to a nonlinear character of the slope efficiency in Figure~\ref{fig:MMLasing}(c) in the  intermediate power regime. Finally, at around \SI{20}{\milli\W}, no new modes are excited and we observe a clear linear behavior with the multi-mode slope efficiency corresponding to 34\%. Considering that the single mode lasing has a slope efficiency of around 5\% (see above), the simultaneous lasing of seven modes is consistent with the observed multi-mode slope efficiency.

\subsection{Comparison with other Ti:sapphire lasers}
One of the very first demonstrations of a low threshold Ti:sapphire laser was made in 1986~\cite{albers1986continuous}, with a threshold of \SI{300}{\milli\watt} and a slope efficiency of 5\% recorded for a mode having a central wavelength of \SI{720}{\nano\meter}, whereas the laser was tunable around a range between \SI{715}{\nano\meter} to \SI{825}{\nano\meter}. 

To classify different Ti:sapphire crystals a figure of merit (FOM) can be defined as the ratio of absorption coefficients for the pump and the emission wavelength, FOM = $\alpha_\text{pump}$/$\alpha_\text{emission}$~\cite{sanchez1988crystal,rapoport_titanium_1988,alfrey1989modeling,pinto1994improved,joyce2010progress,nehari2011ti}. 
Slope efficiencies as high as 42\% and 34\% were demonstrated for high threshold values of \SI{750}{\milli\watt} and \SI{600}{\milli\watt} respectively for Ti:sapphire crystals with different FOMs~\cite{pinto1994improved}. In these experiments the lasing wavelengths range from \SI{665}{\nano\meter} to \SI{1070}{\nano\meter} with pump at \SI{514}{\nano\meter}. 

Furthermore, a low threshold of \SI{90}{\milli\W} with a slope efficiency of 6\% was demonstrated~\cite{harrison1991low} for a mode at \SI{800}{\nano\meter} and with the laser having a tuning range of \SI{775}{\nano\meter} to \SI{829}{\nano\meter}. Other low threshold values reported for Ti:sapphire lasers exist in the range of \SI{120}{\milli\W} to \SI{55}{\milli\W}~\cite{kowalevicz2002ultralow,roth2011direct,wang2016laser,grivas2012tunable,grivas2018generation,roth2009directly}. 

Moreover, the main cavity of choice for Ti:sapphire lasers has been the mirror cavity, with some studies being performed on waveguides and a fibre laser. The most commonly used pump wavelength is \SI{532}{\nano\meter}. All of these values are reported in Table~\ref{tab:literature} and are compared to our results. 
Previously, the lowest recorded threshold values for single-mode lasing was \SI{55}{\milli\watt}~\cite{grivas2018generation} with a slope efficiency of 8.9\% in waveguide based Ti:sapphire laser operating at \SI{798.5}{\nano\meter}. The highest slope efficiency of 42\% was reported for a relatively higher threshold value of \SI{750}{\milli\watt}~\cite{pinto1994improved} for a mode at \SI{800}{\nano\meter}. However, in this case the Ti:sapphire crystal was pumped with multi-mode argon-ion laser source covering a range of \SI{488}{\nano\meter} to \SI{515}{\nano\meter}. Whereas the highest slope efficiency for a Ti:sapphire laser pumped with a single wavelength of \SI{532}{\nano\meter}~\cite{roth2011direct} is 30\% for a threshold of \SI{116}{\milli\watt}. In our experiments we have demonstrated high slope efficiencies for both single-mode and multi-mode operation and the lowest recorded thresholds in both regimes.

\begin{table*}[htbp]
\caption{Lasing threshold and slope efficiency values reported for Ti:sapphire lasers in CW operation. The values given are those reported in the references, though definitions vary slightly.}\label{tab:literature}

\begin{tabularx}{\textwidth}{XXXXXXXX}
\hline\hline
Threshold & Slope & Cavity & $\lambda$ Emission / &  \multirow{2}{*}{$\lambda_p$ (\SI{}{\nano\meter})} & Emission & \multirow{2}{*}{Year} &  \multirow{2}{*}{Reference} \\
(\SI{}{\milli\watt})& efficiency &type & Range (\SI{}{\nano\meter}) & & Regime &&\\
\hline
\rowcolor{Gray}
14.2  & 34\%  & WGMR & 744.4 - 873.5 & 516.6 & Multi-mode & 2021  & This work  \\ 
\rowcolor{LightCyan}
19.5  & 7.7\%  & WGMR & 788.8 & 516.6& Single-mode & 2021  & This work  \\
55\footnote{Found in supplementary information}  & 8.9\%  & Waveguide & 798.5  & 532   & -& 2018 & ~\cite{grivas2018generation}  \\
60  & 15\%  & Mirror & 800  & 532& Multi-mode & 2011 & 
 ~\cite{roth2011direct,kemp2021private}  \\
84  & 4.5\%  & Waveguide & 798.25  & 532& -   & 2012 & ~\cite{grivas2012tunable}  \\
90  & 6\%  & Mirror & 800 & 514.5& -   & 1991 & ~\cite{harrison1991low}  \\
106  & 16\%  & Mirror & 795   & 532& Multi-mode & 2009 & ~\cite{roth2009directly,kemp2021private}  \\
116  & 30\%  & Mirror & 800  & 532& Multi-mode & 2011 & 
 ~\cite{roth2011direct,kemp2021private}  \\
118.2  & 29.6\%  & Fibre Laser & 769 - 832  & 532/520& Multi-mode & 2016 & ~\cite{wang2016laser}  \\
120  & -   & Mirror & 840  & 532& -& 2002 & ~\cite{kowalevicz2002ultralow} \\
220  & 12\% & Mirror & 800 & 488 - 515& - & 1994 & ~\cite{pinto1994improved} \\
300  & 5\% & Mirror & 720 & 488& -&  1986 & ~\cite{albers1986continuous} \\
$\sim$ 400\footnote{Approximated from Figure 3}  & 23.5\%  & Waveguide & 798.25 & 532 & -  & 2012 & ~\cite{grivas2012tunable}  \\
572  & 5\%  & Mirror & 783   & 452 & Multi-mode & 2009 & ~\cite{roth2009directly,kemp2021private}  \\
600  & 34\% & Mirror & 800 & 488 - 515& - & 1994 & ~\cite{pinto1994improved} \\
750  & 42\% & Mirror & 800 & 488 - 515& - & 1994 & ~\cite{pinto1994improved} \\
\hline\hline
\end{tabularx}

\end{table*}

\subsection{Amplification in Ti:sapphire WGL}
\begin{figure*}
  \includegraphics[width=\linewidth]{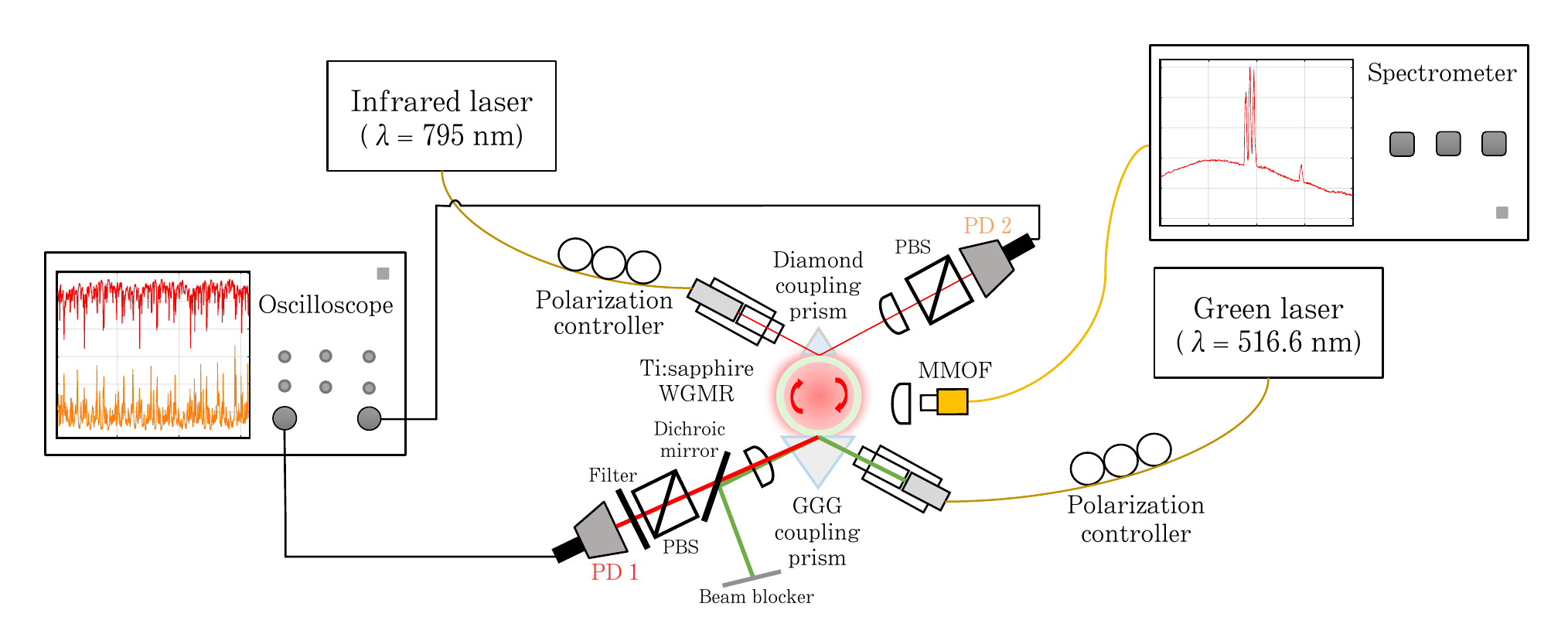}
  \caption{The modified experimental setup. In addition to the GGG prism, a diamond prism is used to couple an infrared laser with the central wavelength of $\lambda=\SI{795}{\nano\m}$. The outputs from both prisms are collected using PDs and are fed to the oscilloscope. Polarization controllers and PBS cubes are used for polarization selection. The spectrometer is used to monitor the lasing excited by the green pump source.}
  \label{fig:Gainsetup}
\end{figure*}

Once the lasing was observed in the Ti:sapphire WGL, the setup was modified, so that we can observe the amplification of a laser probe coupled to the WGM within the gain region, see Figure~\ref{fig:Gainsetup}. A probe laser with a central wavelength of \SI{795}{\nano\m} is coupled in to the resonator using a diamond prism, while the green laser remains coupled through the GGG prism. The resonator rim fluorescence is still collected into the spectrometer, and the excited WGMs are observed with the oscilloscope. For the passive resonator the overall linewidth $\gamma$ of a WGM is given by:
\begin{align}
    \gamma &= \gamma_\text{diamond} + \gamma_\text{int} + \gamma_\text{GGG}, \label{eq:passiveLoss}
\end{align}
where $\gamma_\text{int}$ is the intrinsic linewidth 
due to the material absorption and surface scattering, $\gamma_\text{diamond}$ represents the coupling linewidth of the diamond prism, and $\gamma_\text{GGG}$ shows an additional coupling linewidth due to the GGG prism used to couple the green pump laser. For an active resonator where the green pump laser is turned on and the WGMs excited at \SI{795}{\nano\m}, the gain results in a decrease in the total linewidth:
\begin{align}
    \gamma' &= \gamma - \gamma_\text{gain},\label{eq:activeLoss}
\end{align}
where $\gamma'$ is the linewidth of the same system but with the gain ($\gamma_\text{gain}$) introduced in the WGMR, courtesy of the green pump. 

\begin{figure*}
\begin{tabular}{ c r }
&\multirow{4}{*}{\includegraphics[width=95mm]{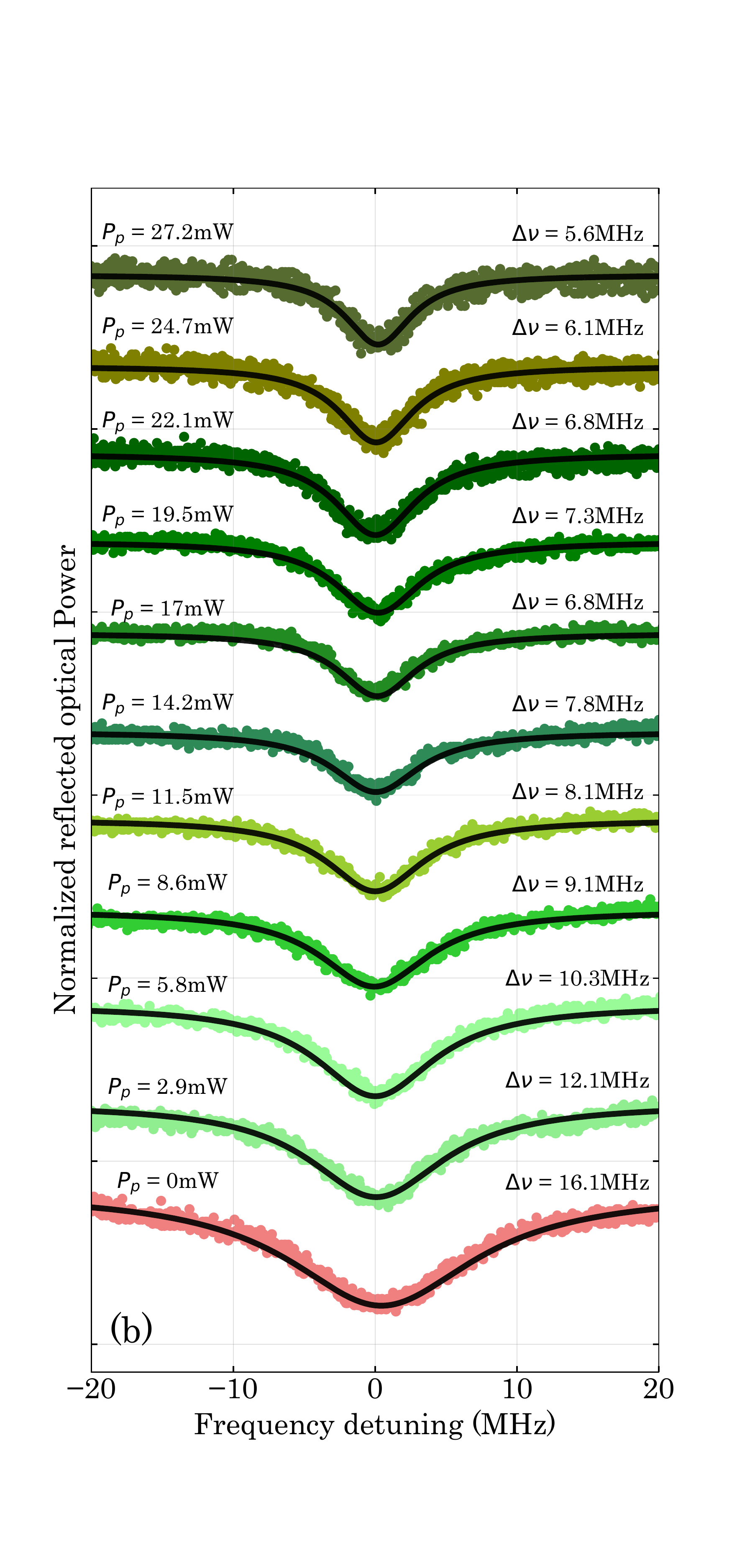}}\\
{\includegraphics[width=85mm]{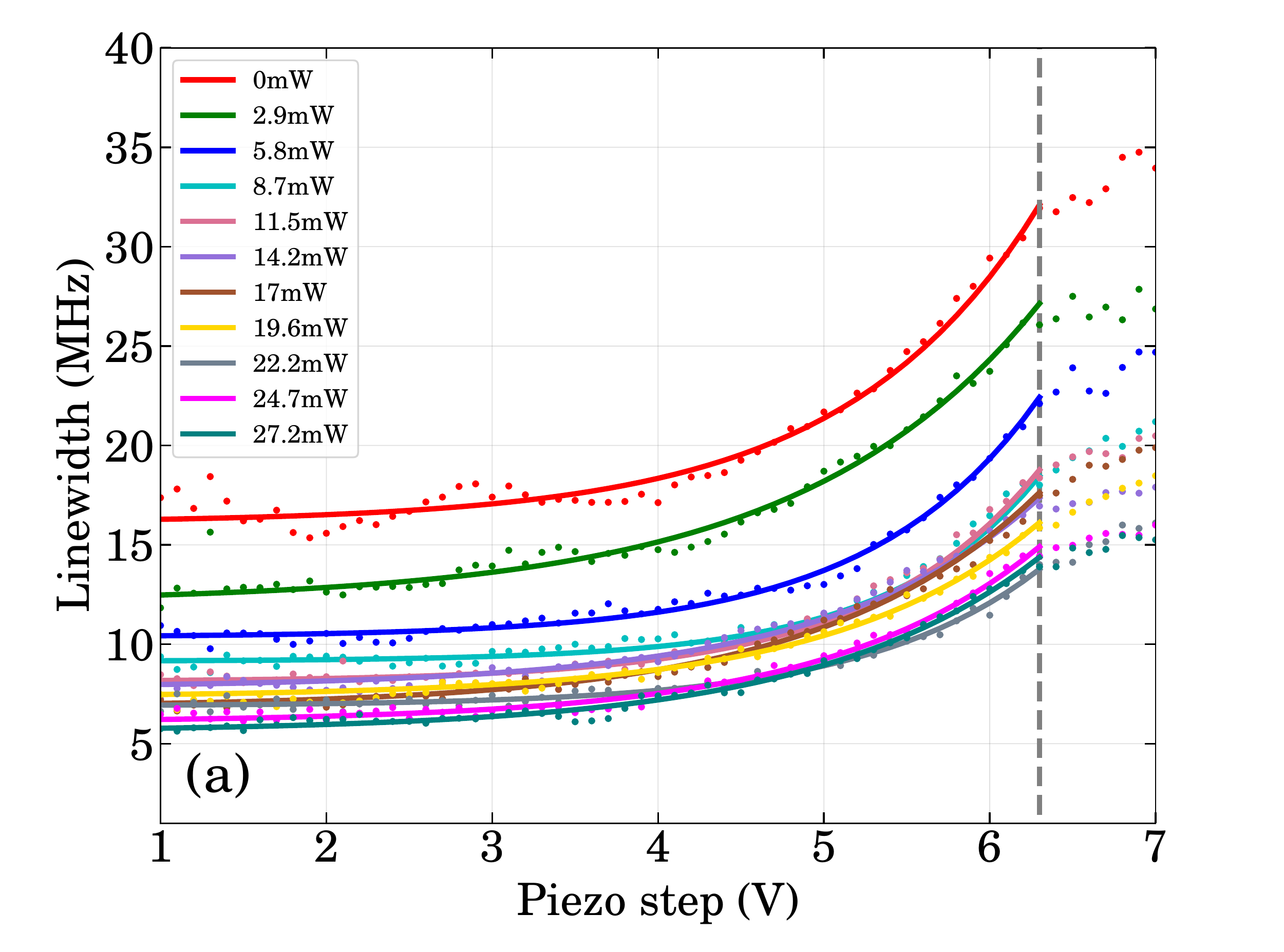}}&\\
{\includegraphics[width=60mm]{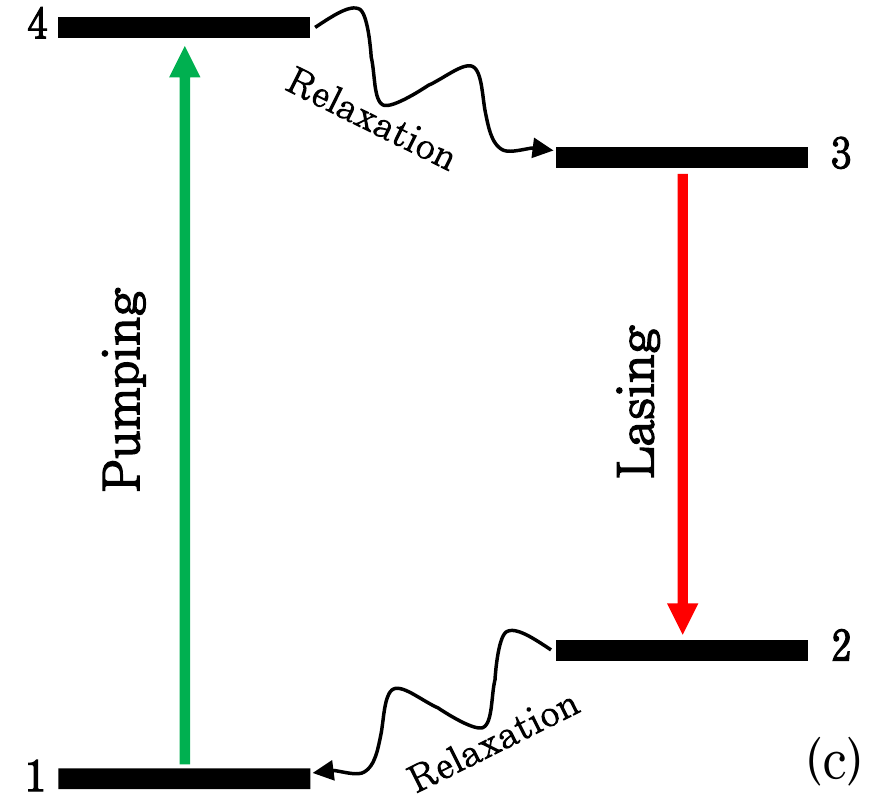}}&\\
{\includegraphics[width=85mm]{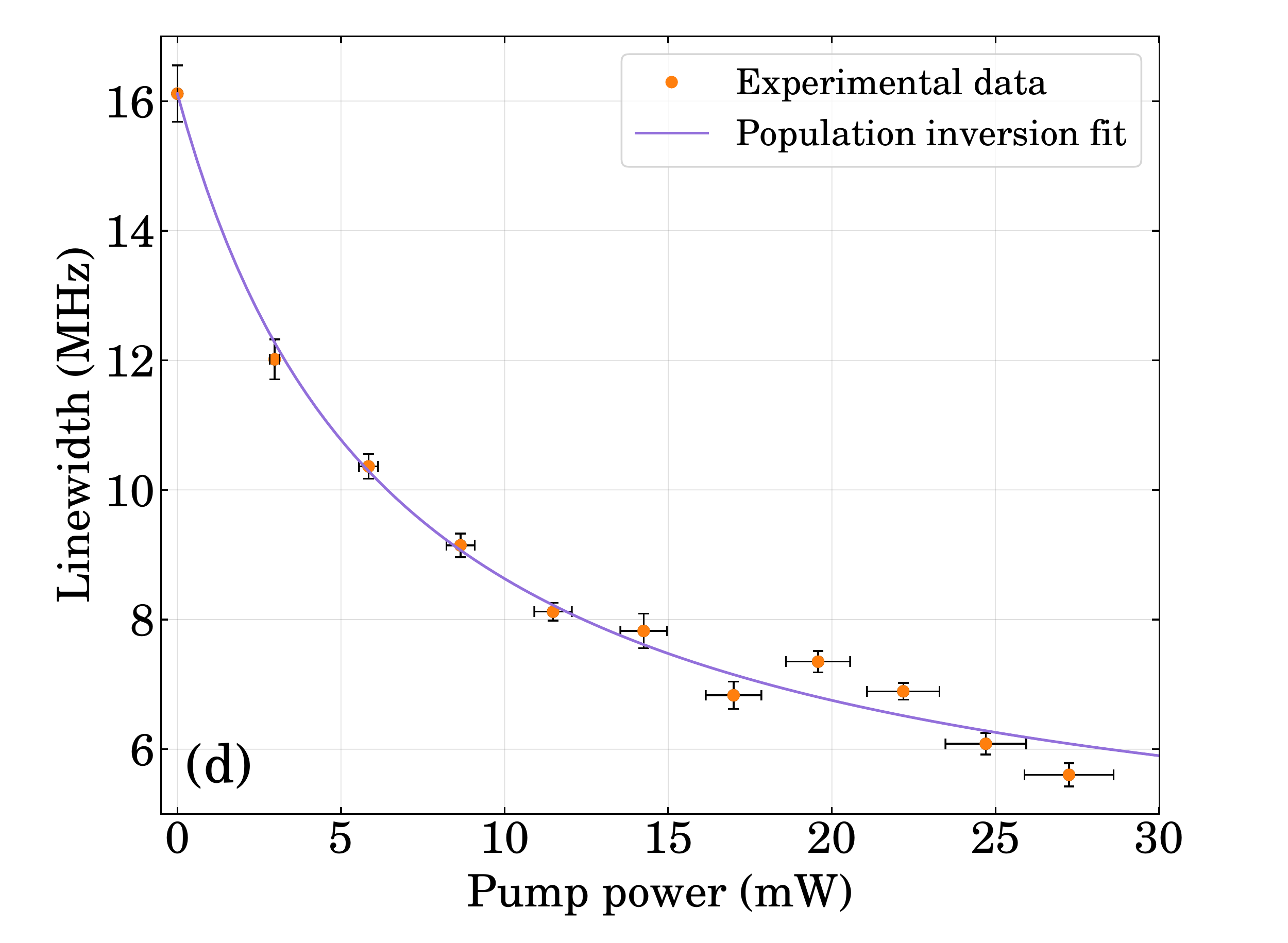}}&\\ 
\end{tabular}
\caption{Linewidths measured using the \SI{795}{\nano\m} laser while increasing the pump power. (a) Shows the linewidth of the same high $Q$-factor \SI{795}{\nano\m} mode versus the distance between the resonator and the prism. Each point is the result of a Lorentzian fit to the resonance's line shape, such as shown in subplot (b). The distance between the resonator and the prism decreases when the voltage of the piezo is increased. The dashed line indicates the touching point for the prism and the WGMR. Exponential fits are applied to all of the acquired data sets to the left of the touching point for different pump powers. This allows us to track linewidth changes caused by the green pump. (b) Shows the linewidth of the mode at a fixed distance and at different pump powers. The linewidth of the probe laser mode is \SI{16.1}{\mega\Hz} when the pump laser is off and decreases down to \SI{5.6}{\mega\Hz} at the maximum pump power, i.e., \SI{27.2}{\milli\W}. The linewidths are measured by applying Lorentzian fits to the data shown in black. (c) Shows the energy diagram for a four-level laser system. Ti:sapphire being a four-level laser material, lases between level 3 and 2. (d) The linewidth of the mode extracted in Figure~\ref{fig:LWGain}(a) in the limit of no coupling to the \SI{795}{\nano\meter} mode with the error margins of the fits on the y-axis and $\pm5\%$ accuracy correction of the Thorlabs power meter on the x-axis. A fit is applied to the data showing the population inversion of a four-level laser system shown in Figure~\ref{fig:LWGain}(c).}\label{fig:LWGain}
\end{figure*}

In order to show the linewidth narrowing expressed by equation~(\ref{eq:activeLoss}), we first choose a WGM which is excited by the probe laser at approximately \SI{795}{\nano\m}. Then the distance between the GGG (pump) prism and the WGMR is fixed such that it provides the maximum gain, which is confirmed by observing lasing with the spectrometer. Next we record the linewidth values of a high $Q$-factor mode while moving the diamond prism using the piezo positioner (see figure S7 in section 6 of the supplemental document), from far away (undercoupled regime) to all the way to touching the resonator (maximally overcoupled regime), at various powers of the green pump, see Figure~\ref{fig:LWGain}(a). The dashed grey vertical line in the figure at \SI{6.3}{\volt} shows the position where the diamond prism is touching the resonator. In each case, an exponential has been fitted to the data before the resonator and prism touch. The asymptote of this exponential linewidth dependence gives the linewidth in the limit that there is no coupling to the diamond prism, i.e., $\gamma^\prime=\gamma_\text{int}+\gamma_\text{GGG}-\gamma_\text{gain}$. Here the measured gain is the so called small-signal gain, where any saturation effects are ignored, as we are in the limit that no light of \SI{795}{\nano\m} is coupled into the resonator. These asymptotic linewidth values for a fixed distance (voltage) are plotted in Figure~\ref{fig:LWGain}(b).

The theoretical fit to the linewidth vs.\ pump power is found from the expression for the gain in a four-level laser systems with titanium number density $\rho_\mathrm{Ti}$ and population number densities $\rho_\mathrm{Ti}=\rho_1+\rho_2+\rho_3+\rho_4$, see Figure~\ref{fig:LWGain}(c)~\cite{saleh_fundamentals_2007}
\begin{equation}
\gamma_\text{gain}=\eta_\text{olp} \frac{c}{n_l}\sigma_{23}\rho_\text{Ti}\frac{\Delta \rho}{\rho_\mathrm{Ti}}, \label{eq:GammaGain}   
\end{equation}
where $\eta_\text{olp}<1$ is a parameter describing the transverse overlap between the pumping and lasing WGMs, $n_l$ the refractive index of the lasing mode, and $\sigma_{23}$ is the emission cross section. For perfectly overlapping modes $\eta_\text{olp}=1$, but in realistic scenarios $\eta_\text{olp}<1$. In our proof-of-principle experimental setup we do not have a control over $\eta_\text{olp}$, and will consider it as a free parameter. The population inversion of a four-level laser system is given by~\cite{saleh_fundamentals_2007}
\begin{align}
    \frac{\Delta \rho}{\rho_\mathrm{Ti}}=\frac{\rho_3 - \rho_2}{\rho_\mathrm{Ti}} &\approx \frac{W_p \tau_\text{rad}}{1 + W_p \tau_\text{rad} + W_l \tau_\text{rad}},    \label{eq:popinversion}
\end{align}
where $\tau_\text{rad}=\tau_{3}\approx \SI{3.2}{\micro\second}$ is the life time of the excited state of the lasing transition, and the pumping (emission) rates $W_p$ ($W_l$) are related to the intracavity intensities $I_p$ ($I_l$) as
\begin{align}
    W_p = I_p\frac{\sigma_{14}}{\hbar\omega_p},\quad\text{and}\quad W_l = I_l\frac{\sigma_{23}}{\hbar\omega_l}
\end{align}
with the pump (emission) frequency $\omega_p$ ($\omega_l$), the absorption cross section of the pump $\sigma_{14}$, and the emission cross section $\sigma_{23}$. 

The intracavity pump intensity can be estimated as 
\begin{equation}
I_p=2 \pi R\frac{P_{p}}{V_p}\frac{\mathcal{F}_p}{\pi},     
\end{equation}
where $V_p$ is the pump mode volume, hence $V_p/(2\pi R)$ is its cross section, where $R$ is the resonator's radius. The mode volume is estimated in Section 2.1 of the supplemental document. The factor $\mathcal{F}_p/\pi$ is the intracavity power enhancement of the pump assuming that the mode is approximately critically coupled, where 
\begin{equation}
    \mathcal{F}_p=\frac{\lambda_pQ_p}{2\pi Rn_p}
\end{equation}
is the pump mode finesse, with the pump wavelength $\lambda_p$, the pump quality factor $Q_p$, and the refractive index at the pump frequency $n_p$.

Typically $W_l$ can be ignored in Equation \eqref{eq:popinversion}. The mode that we measured, however, did not lase at high pump power. We therefore know that there were competing modes with better pump overlaps so $W_l$ is not necessarily negligible in our experiment, as the light emitted into other modes that overlap with our chosen mode contributes to it.
We can estimate $W_l$ by reasoning that the photons at the fluorescence wavelength have been created by emission after a pump photon has been absorbed.
It is well observed that the spontaneous emission of a laser below threshold grows proportional to its pump power.
The intensity of fluorescence is then
\begin{equation}
    I_l = \eta_\text{spon}I_p\frac{ \omega_l}{\omega_p}\frac{\mathcal{F}_l}{\mathcal{F}_p},
\end{equation}
where $\eta_\text{spon}<1$ is the fraction of pump photons that undergo spontaneous emission $(3\rightarrow2)$ below threshold. The other factors account for the energy difference of the photons and the different intracavity field enhancement.

Now we have the fitting function of a form
\begin{equation}
    \gamma^\prime(P_p)=\gamma-\frac{AP_p}{B+P_p},
\end{equation}
where 
\begin{equation}
A = \eta_\text{olp}\frac{c}{n_l}\sigma_{23}\rho_\mathrm{Ti},\quad{\rm and}\quad B = \frac{\hbar\omega_p V_p}{2 R\tau_\text{rad}(\sigma_{14}F_p + \eta_\text{spon}\sigma_{23}F_l)}.
\end{equation}
This fit is shown as the solid line in Figure~\ref{fig:LWGain}(d). It is in excellent agreement with the data. The error bars originate from the Lorentzian fit on $y$-axis and the power meter on the $x$-axis.
From the fitted values of $A$ and $B$ we can estimate that
\begin{equation}
    \eta_\text{olp}\approx\num{3.3e-4},\quad\text{and}\quad\eta_\text{spon} \approx 0.074.
\end{equation}
From the magnitude of $\eta_\text{spon}$, we see that intracavity spontaneous emission is indeed significant during this experiment.
The value of $\eta_\text{olp}^{-1}$ roughly gives the magnitude of the gain improvement that is theoretically possible by facilitating a better overlap of the modes e.g., by optimizing the rim shape, or simply by choosing a better overlapping pair of the pump and amplified modes available in the present resonator geometry.

\section{Conclusion}
In conclusion, we have for the first time demonstrated a Ti:sapphire WGL. The WGL operates in both single-mode and multi-mode regimes, with a range spanning over \SI{100}{\nano\m} and a high slope efficiency of 34\%. The threshold value of \SI{14.2}{\milli\W} reported, is the lowest reported for a Ti:sapphire laser to the best of our knowledge. Moreover, we demonstrated a novel method of using the gain in the resonator to enhance the $Q$-factor of WGMs at \SI{795}{\nano\m}, due to the gain inside the resonator acting as a negative loss mechanism for a wavelength within the excitation range of Ti:sapphire. The trend shown by the linewidth values of the \SI{795}{\nano\m} WGM with the increase of the pump power was explained using the population inversion model of a four-level laser system. 

There are a number of opportunities for further investigations of this unique solid-state laser platform. By reducing the mode volume in the resonator the threshold should drop even further. Selectively pumping the region of the rim of the resonator through Bessel beams~\cite{schwefel_directionality_2004} would reduce the losses induced by the green coupling prism~\cite{savchenkov2006optical,ilchenko2007efficient} at the cost of losing the resonant enhancement of the green mode (finesse $\sim11$). This will also provide more control over the pump light overlap with the lasing modes, and make the pump intensity independent from the lasing modes coupling rate. The pump laser used in this study could only achieve a highest absorbed pump power of \SI{27.2}{\milli\W}; at higher pump powers the range of this Ti:sapphire WGL will increase. Exploring Ti:sapphire crystals of different FOM to fabricate the WGMR can also result in better threshold and slope efficiencies. As it has been shown experimentally before that different FOM crystals can result in better outputs, i.e., lower thresholds, higher slope efficiencies and higher output powers~\cite{pinto1994improved,roth2009directly}.
Finally, the experimental results reported in this work are very promising for the implementation of a more compact and efficient Ti:sapphire laser.

\section{Experimental methods}\label{sec:methods}
\subsection{WGMR fabrication and testing}\label{sec:methodsFabrication}
At the core of this study is the Ti:sapphire WGMR. First, a piece was cut from a Ti:sapphire crystal (titanium density $N=\SI{7.3+-5e24}{\per\meter\cubed}$, see Section 1 of the supplemental document) with a diamond saw. The crystal was cut in the $z$-cut configuration, i.e., the optic axis of the crystal coincides with the symmetry axis of the resonator. Afterwards, the $z$-cut piece was glued to a brass rod using mounting wax, as shown in Figure~\ref{fig:Fabrication}(a). Next the curvature of the resonator surface was shaped using a dental drill, while the brass rod was rotating via a turning machine. The curved disc was then polished using basic diamond slurry solutions of $\text{pH}\approx10$ with progressively smaller particle sizes. The particle size was chosen depending on the coarseness of the surface of the resonator. First the resonator was polished using polycrystalline diamond slurry with \SI{30}{\micro\meter} average particle size; the quality of the resonator surface after this round of polishing is shown in Figure~\ref{fig:Fabrication}(b).
The resonator was further polished with \SIlist{9;3;1;0.25}{\micro\meter} slurries. The results after polishing with \SIlist{9;3}{\micro\meter} are shown in Figure~\ref{fig:Fabrication}(c) and (d). At each stage of polishing the surface roughness is improved. After each step of polishing the resonator is cleaned thoroughly to avoid cross-contamination of diamond particles.

The final resonator can be seen in Figure~\ref{fig:Fabrication}(e) and (f). The side-view of the finalized Ti:sapphire WGMR, captured using a microscope can be seen in Figure~\ref{fig:Fabrication}(e). The resonator has a thickness of $\SI{0.78}{\milli\m}$ and a radius of curvature of $r = \SI{0.8}{\milli\m}$. The top-view of the resonator showing it's radius, $R = \SI{2.175}{\milli\m}$ can be see in Figure~\ref{fig:Fabrication}(f). The $Q$-factor of the resonator was tested throughout the polishing process. The best recorded values for the critically coupled linewidths at \SI{795}{\nano\m} and \SI{1550}{\nano\m} are \SI{7.80}{\mega\Hz} and \SI{1.93}{\mega\Hz} respectively, as shown in Figure~\ref{fig:Fabrication}(g). These linewidth values amount to critical $Q$-factors of \num{5e7} and \num{1e8} and intrinsic $Q$-factors of \num{1e8} and \num{2e8} at \SI{795}{\nano\m} and at \SI{1550}{\nano\m} respectively. These are the highest $Q$-factor values reported for a Ti:sapphire WGMR at the aforementioned wavelengths to the best of our knowledge.

\subsection{Lasing setup}\label{sec:LasingSetup}
 
The main component of our experiment is the Ti:sapphire WGMR. After the fabrication process, the resonator was transferred to the experimental setup depicted in Figure~\ref{fig:Lasingsetup}.  A green laser ($\lambda = \SI{516.6}{\nano\m}$) is evanescently coupled into the resonator using a Gadolinium Gallium Garnet (GGG) prism ($n$ = 1.96)~\cite{wood_optical_1990}. The focus of the green laser is controlled using a graded index (GRIN) lens and a pigtailed ferrule. The distance between the prism and the resonator can be controlled using a piezo positioner with nanometer accuracy. Due to sapphire's minimal birefringence, light can be coupled into both polarizations at the same angle, which makes it very easy to switch between the TE and TM polarizations using a three paddle polarization controller (PC), without realigning the beam. The output light from the prism is collimated using a lens and is split by a dichroic mirror into the green pump light and the fluorescence/lasing light. The green light is collected by a silicon (Si) photodiode (PD), whereas the lasing light is coupled into a single mode optical fibre (SMOF), via a collimating lens, aspheric collimator and two adjustable mirrors. A polarizing beam splitter (PBS) is placed right before the PD to confirm the polarization of the WGM. The PD is connected to an oscilloscope to monitor the WGMs. The output from the SMOF is coupled into an optical spectrum analyzer. Another Si PD is used to collect the surface scattered WGMR light to observe the WGM lasing. An optical filter is placed before this PD to make sure that it only collects the lasing light while filtering the scattered green light. A spectrometer together with a multi mode optical fibre (MMOF) is also used to observe the fluorescence spectrum of the resonator.

The pump powers reported in this work are the powers coupled into the resonator. The pigtailed ferrule utilized in the experiment is not designed for our pump wavelength, so it causes extra loss. Additionally, neither of the pigtailed ferrule, GRIN lens or GGG prism are coated. As the loss from these elements could be mitigated by specially designed replacements, we use the power at the resonator input (i.e.,\ after one reflection of the prism). This power is calculated assuming the losses at the input and output of the prism are equal, as they are for Fresnel reflections. Imperfect mode matching of the pump beam to the pump whispering-gallery mode also reduces the power that enters the resonator. As an optimized prism coupling setup can achieve better than \SI{99}{\percent} coupling efficiency~\cite{strekalov2009efficient}, we have further corrected the power to be the power that actually enters the resonator. See the Supplement for further details on the powers used.

\subsection{Amplifier setup}\label{sec:methodAmplifier}
After observing lasing in the Ti:sapphire resonator, we modified the experiment to observe amplification inside the WGMR stimulated by the pump and measure the gain. This modified setup is depicted in Figure~\ref{fig:Gainsetup}. In order to observe the gain, we used another prism made from diamond ($n=2.38$~\cite{phillip_kramers-kronig_1964}), together with the previously used GGG prism. The diamond prism is used to couple an infrared (IR) probe laser with a central wavelength of \SI{795}{\nano\m} to one of the lasing  WGMs. This wavelength lies within the gain range of the Ti:sapphire emission when it is pumped by a green laser.

Once the \SI{795}{\nano\m} probe laser is coupled into the resonator and WGMs are observed, the \SI{516.6}{\nano\m} pump laser is turned on to observe the probe amplification. Polarization controllers after both the green and the infrared lasers are used to switch between TE and TM polarizations. The position of the GGG prism is fixed during the process of data collection, whereas the diamond prism is moved away or towards the resonator, as the linewidth of the probe WGM is measured and recorded. The spectrometer is used to monitor the lasing excited by the green pump source. Two PDs are used to collect the output light from both prisms, with a PBS in front of each to monitor the polarization of the modes. The detectors are connected to the oscilloscope and are used to observe the resonances excited by the \SI{795}{\nano\m} probe laser.

\medskip
\textbf{Supporting Information} \par 

Supporting Information is available.

\medskip

\textbf{Acknowledgements} \par 

We would like to thank Mathew Denys. We would like to acknowledge fruitful discussions with Jevon Longdell. We would like to thank Joseph Borbely and the Measurement Standards Laboratory of New Zealand (MSL). We would like to thank Brent Pooley and Malcolm Reid for their help in determining the titanium dopant concentration. AG would like to thank the Chinese National Science Fund for Talent Training in the Basic Sciences (No.~J1103208) for sponsoring his visit to Otago. DVS and HGLS would like to thank the Ministry of Business, Innovation and Employment Catalyst Leaders fellowship (20-UOO-001-ILF).




%

\end{document}


\maketitle

\tableofcontents{}

\begin{abstract}
Titanium doped sapphire (Ti:sapphire) is a laser gain material with broad gain bandwidth benefiting from the material stability of sapphire. These favorable characteristics of Ti:sapphire have given rise to femtosecond lasers and optical frequency combs. Shaping a single Ti:sapphire crystal into a millimeter sized high quality whispering gallery mode resonator ($Q\sim10^8$) reduces the lasing threshold to \SI{14.2}{\milli\W} and increases the laser slope efficiency to 34\%. The observed lasing can be both multi-mode and single-mode. This is the first demonstration of a Ti:sapphire whispering gallery laser. 
Furthermore, a novel method of evaluating the gain in Ti:sapphire in the near infrared region is demonstrated by introducing a probe laser with a central wavelength of \SI{795}{\nano\m}. This method results in decreasing linewidth of the modes excited with the probe laser, consequently increasing their $Q$. These findings open avenues for the usage of whispering gallery mode resonators as cavities for the implementation of compact Ti:sapphire lasers. Moreover, Ti:sapphire can also be utilized as an amplifier inside its gain bandwidth by implementing a pump-probe configuration.
\end{abstract}

\section{Titanium concentration measurement}
\begin{figure}
\centering
  \includegraphics[width=0.6\linewidth]{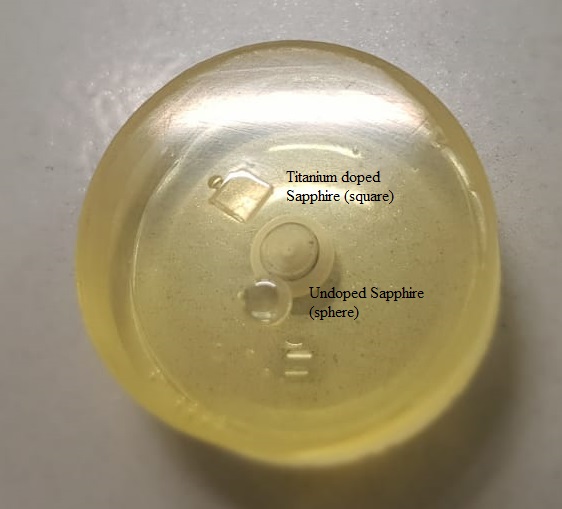}
  \caption{A chunk of the Ti:sapphire crystal used to fabricate our WGMR used in our experiments is shown in square and an undoped sapphire sphere mounted using epoxy can be seen in yellow.}
  \label{fig:concentration}
\end{figure}

In order to quantify the concentration of the titanium dopant in the Ti:sapphire crystal, an undoped sapphire chunk was used as reference. Figure~\ref{fig:concentration} shows a chunk from the Ti:sapphire crystal (square shaped) and an undoped sapphire sphere, mounted using epoxy. The mounted samples were then placed into inductively coupled plasma mass spectrometry (ICP-MS) setup and the concentration results for the titanium doped and the undoped sapphire were compared. The acquired results showed concentrations of titanium in the doped sapphire of \SI{145+-100}{\micro\gram\per\gram}. By using the following conversion we found the concentration of titanium dopant in our Ti:sapphire crystal.

\begin{align}
    \rho_\mathrm{Ti} 
    &= \frac{N_\text{Ti}}{V} = \frac{m_\text{Ti}}{m_\text{sapphire}} \frac{m_\text{sapphire}}{V} \frac{N_\text{Ti}}{m_\text{Ti}},\nonumber
    \\
    &= \text{(measured mass ratio)} \times \text{(sapphire density)} \times \frac{N_\text{A}}{\text{molar mass of Ti}},\nonumber
    \\
    &= \num{145+-100e-6}  \times \SI{3.98e6}{\gram\per\meter\cubed} \times \frac{\SI{6.022e23}{\per\mol}}{\SI{47.867}{\gram\per\mol}},\nonumber
    \\
    &= \SI{7.3+-5.0e24}{\per\meter\cubed}.  \label{eq:concentrateTI}
\end{align}

The calculated value shown in equation~(\ref{eq:concentrateTI}) is close to the typical titanium doping concentration in a Ti:sapphire crystal~\cite{renk2012basics}, i.e., \SI{1e25}{\per\meter\cubed}.

\section{Linewidth of WGMs excited by pump laser}
\begin{figure}[hbt!]
\centering
  \includegraphics[width=0.8\linewidth]{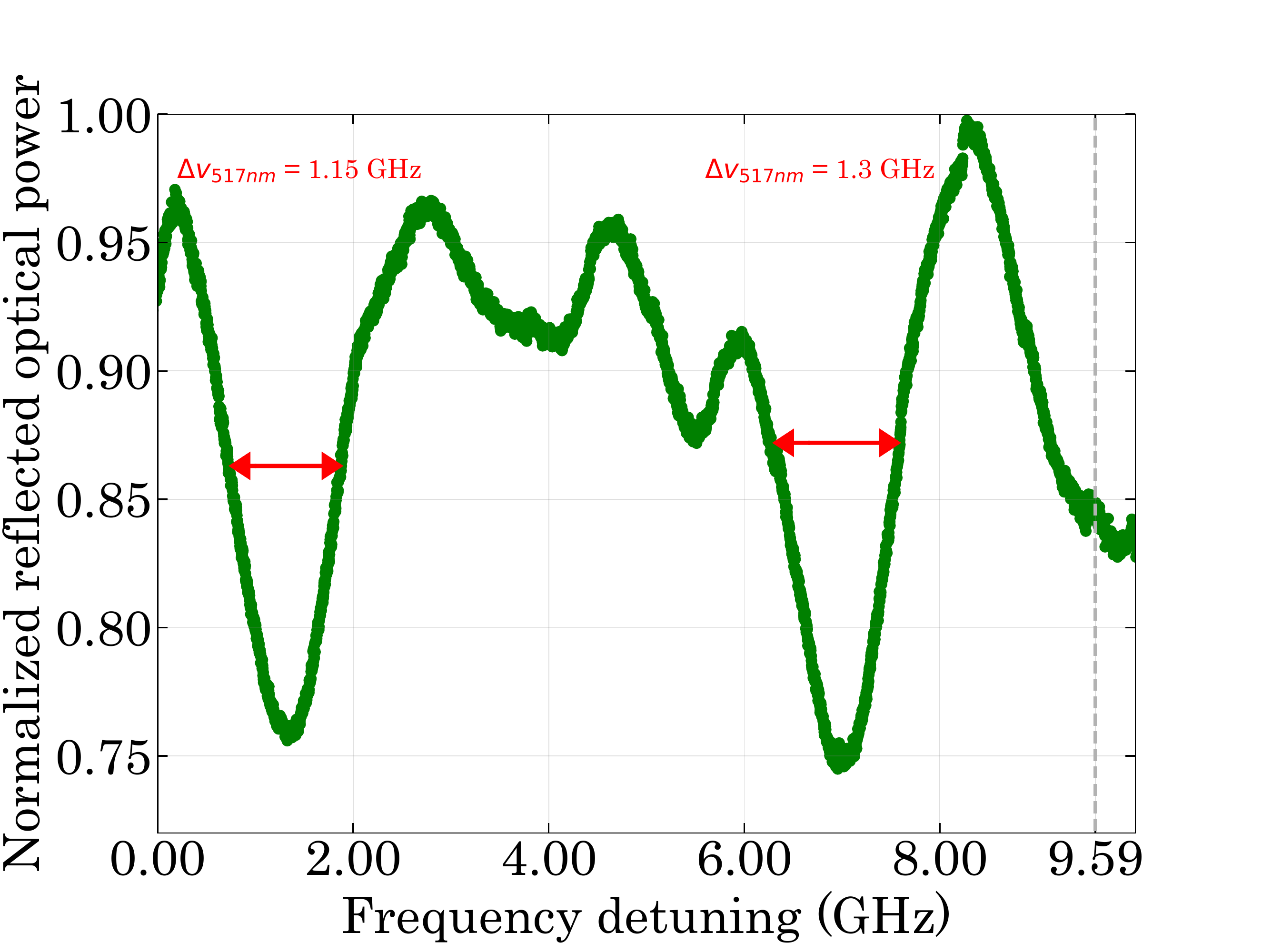}
  \caption{The linewidth of TE-polarized WGMs excited by the pump laser. Normalized reflected optical power versus frequency detuning (GHz). The observed linewidths for the two well defined modes are  \SI{1.15}{\giga\Hz} and \SI{1.3}{\giga\Hz}}
  \label{fig:greenlinewidth}
\end{figure}

Figure~\ref{fig:greenlinewidth} shows the linewidth of a WGM excited by the green pump laser in the lasing setup. The data was collected using a photodiode and recorded through an oscilloscope. The measured linewidth is $\sim\SI{1.15}{\giga\Hz}$, which amounts to a $Q$-factor of $5\times10^5$. This $Q$-factor at a green wavelength in a Ti:sapphire WGMR is the highest recorded to the best of our knowledge.

We can use this measurement to calculate a lower bound on our crystal's figure of merit (FOM). It is
\begin{equation}
    \text{FOM} = \frac{\alpha_{517}}{\alpha_{795}} \geq \frac{\gamma_{i,517}}{\gamma_{i,795}} \geq \frac{\SI{575}{\mega\hertz}}{\SI{3.9}{\mega\hertz}} \sim 150,
\end{equation}
where $\alpha_{517}$ ($\alpha_{795}$) is the absorption coefficient at \SI{517}{\nano\meter} (\SI{795}{\nano\meter}), and $\gamma_{i,517}$ ($\gamma_{i,795}$) is the intrinsic linewidth at \SI{517}{\nano\meter} (\SI{795}{\nano\meter}). It is a lower bound as the modes at \SI{795}{\nano\meter} could be scattering limited, and we have assumed the pump was critically coupled, when in fact it was undercoupled to an unknown degree. Additionally the minor difference in group index at the two wavelengths will increase this number slightly.

\section{Calculation of volume of WGMs}
\begin{figure}[t!]
\centering
\includegraphics[width=0.9\linewidth]{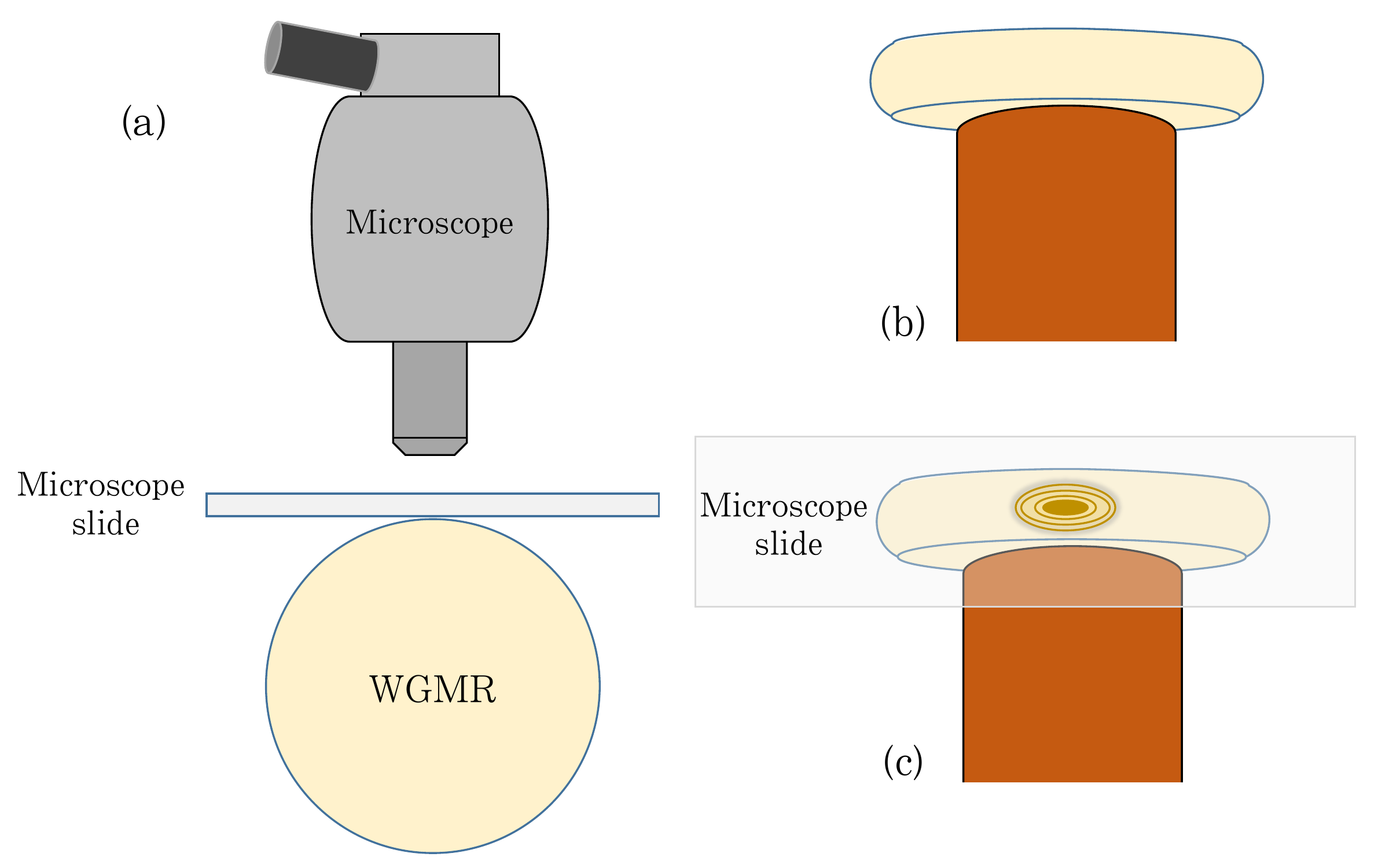}
\caption{Method used to find the volume of modes ($V_p$) of the Ti:sapphire WGMR. (a) shows the side-view of the setup. A microscope glass slide is placed on the equator of the WGMR to form fringes due to interference observed using the microscope. (b) and (c) show the top-view of the setup. (b) shows the resonator without the glass slide, whereas (c) shows the resonator with the glass slide placed on top of it, with the interference pattern forming in the curvature of the resonator.}\label{fig:fringe}
\end{figure}

The volume of the WGMs of the Ti:sapphire WGMR is measured using the method depicted in Figure~\ref{fig:fringe} and explained in~\cite{strekalov2009efficient,sedlmeir2016crystalline}. Figure~\ref{fig:fringe}(a) shows the side view of the setup, a thin microscope glass slide is placed on top of the WGMR, the resulting reflection gives rise to interference fringes shown in Figure~\ref{fig:fringe}(c).

\begin{figure}[t!]
\centering
\includegraphics[width=0.5\linewidth]{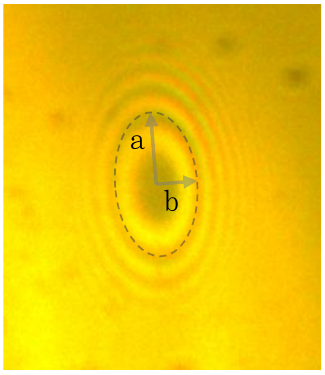}
\caption{The Newton's rings formed when a glass slide is placed on the side surface of the resonator, i.e., the mode volume.}\label{fig:Newtonsrings}
\end{figure}

The experimentally observed image of the Newton's rings which are formed when a glass slide is placed on the top of the WGMR mode volume as shown in Figure~\ref{fig:Newtonsrings}. Where a and b depict the two axes of the elliptical shaped pattern formed due to interference and are given by the following equations:

\begin{align}
    \text{$a$} &= \sqrt{j \lambda R}
    \
\end{align}

\begin{align}
    \text{$b$} &= \sqrt{j \lambda r}
    \
\end{align}
where $j$ is the number of fringes formed due to the interference. Next the ratio between the radii is found using the following equation:

\begin{align}
    \frac{R}{r} &= \frac{a^2}{b^2}
    \
\end{align}
where the radius of curvature of the resonator is given by $r$. This gives us a precise value for $r$ of the resonator, as the $r$ undergoes changes during the fabrication process, i.e., it differs at the time of cutting and after the polishing process. Afterwards, the volume of WGMs was calculated using the following equations:

\begin{equation}
    m \approx \frac{2 \pi R n}{\lambda},\quad\text{when }m\gg 1
\end{equation}
\begin{equation}
    V_p \approx 15.12 R^\frac{11}{4} r^\frac{1}{4} m^{-\frac{7}{6}} \label{eq:volumemode}
\end{equation}
Background and derivation of equation~(\ref{eq:volumemode}) can be found in~\cite{demchenko2013analytical}.

\section{Power measurement}
\begin{figure}[h]
\centering
  \includegraphics[width=1\linewidth]{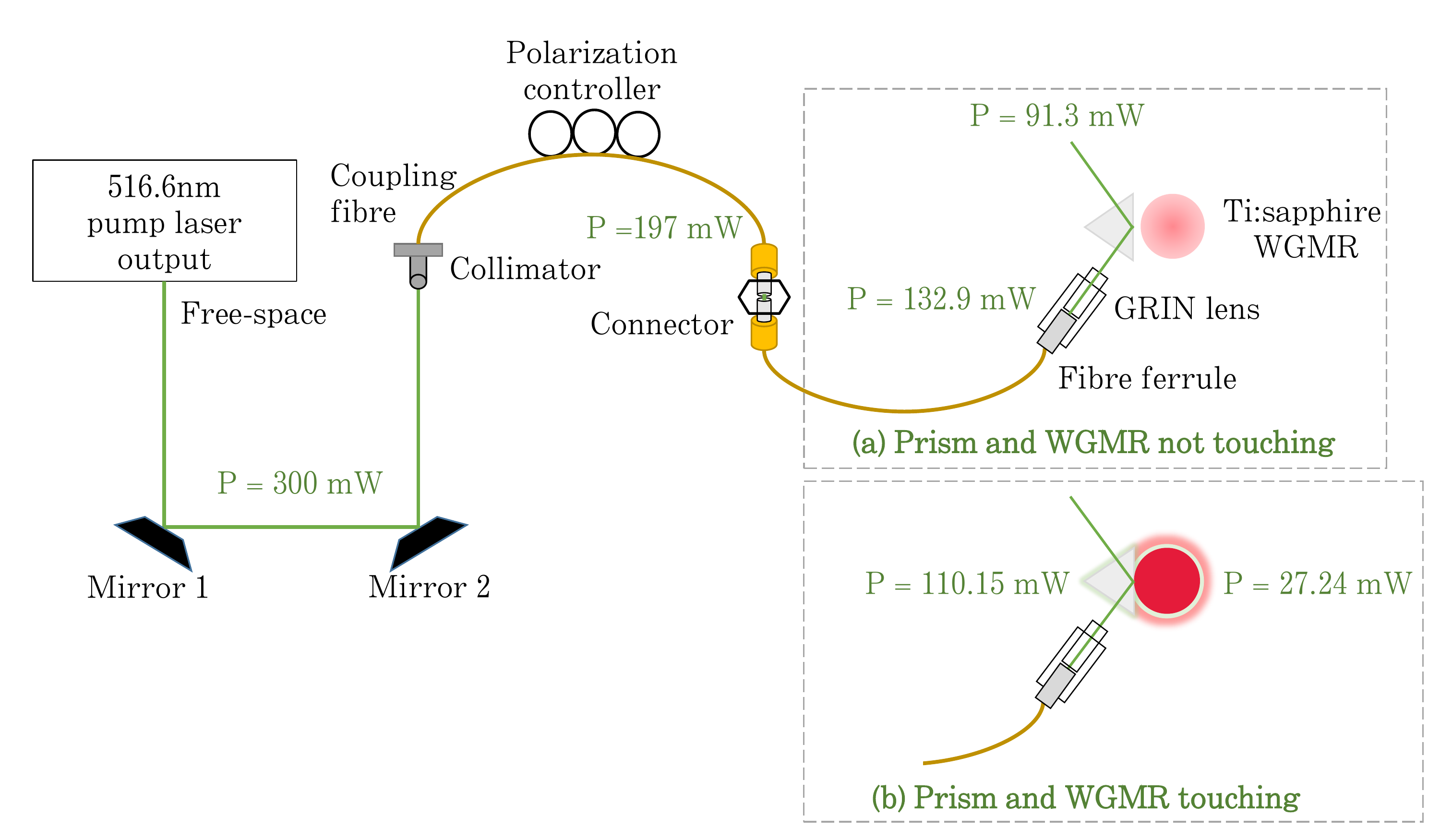}
  \caption{Schematic showing the power measured at different points in the experiment. The highest achievable power with our pump laser is \SI{300}{\milli\watt} in free-space. The free-space output is coupled into a single mode optical fibre and the highest measured power at the output end of the fibre is \SI{197}{\milli\watt}. The fibre is then connected to a pigtailed ferrule fibre focused with the help of a GRIN lens. The output at the end of the ferrule-GRIN lens is \SI{132.9}{\milli\watt}. This drop in power can be attributed to the un-coated GRIN lens. The output from the GRIN lens is incident on the GGG coupling prism. The coupling prism is also un-coated and takes away a chunk of the power. The power measured at the second prism surface is \SI{91.3}{\milli\watt}, this means that the power inside the prism is \SI{110.15}{\milli\watt} and when the pump is coupled into the WGMR only 25\% of the light is coupled. This brings the final power value, i.e., incoupled pump power to \SI{27.24}{\milli\watt}. }
  \label{fig:pumpmeasure}
\end{figure}

The power of the green pump laser was monitored at different points in the experimental setup, using a power meter. Figure~\ref{fig:pumpmeasure} depicts the setup and the process. The pump laser was operated at its maximum output power, i.e., \SI{300}{\milli\W}. The direct output from the laser is in free-space, we used a single mode optical fibre to couple the laser using the beam walking method via two mirrors and an aspheric collimator. We were able to achieve 65\% coupling, i.e., we were able to get an output of \SI{197}{\milli\W} from the incoming \SI{300}{\milli\W} in free-space. Next a three paddle polarization controller was used to change the polarization of the light. The fibre was then connected to another fibre with a pigtailed ferrule output. A GRIN lens was used to focus the light from the ferrule and the pump power was measured at this point to be \SI{132.9}{\milli\W}.
The polarization controller was rotated to get TE-polarized light.
The output from the fibre ferrule-GRIN lens combination was passed through the GGG coupling prism and was measured to be \SI{91.3}{\milli\W}, which amounts to a total loss of about 31\% resulting due to the prism, when the prism is not touching the WGMR. It can be assumed that each face of the prism contributes equally and thus adds about 17\% of loss, so the power inside the prism will be approximately \SI{110.15}{\milli\W}.
The expected loss due to Fresnel reflections for the TE polarization at a GGG--air interface at the coupling angles used is 16.5\%.
Next the coupling efficiency of the WGMs was measured through data acquired from oscilloscope, plotted in Figure~\ref{fig:greenlinewidth}. A coupling efficiency of 25\% was observed, i.e., 25\% of the pump light is getting absorbed into the WGMR. The baseline of Figure~\ref{fig:greenlinewidth} dropped when the prism was moved close to the resonator. We attribute this effect to coupling into low $Q$, deep inlying modes which may contribute to lasing, but due to the small overlap of these modes their overall contribution is negligible. 

Finally the lasing power was measured at the output of the prism. All of the power values measured at various points in this setup are reported in Table~\ref{tab:PumpPower}. The pump was varied from \SI{30}{\milli\W} to \SI{300}{\milli\W} with an incremental step of \SI{30}{\milli\W} in free-space to record these power values.

\begin{table}[hbt!]
\caption {Input pump power ($P_p$) and Lasing power measurement. All powers are measured in milliwatt~(mW).}\label{tab:PumpPower}
\begin{center}
\begin{tabular}{|c|c|c|c|c|c|}    
\hline
Free-space& Fibre coupled &$P_p$ after & $P_p$ after & Incoupled  &Lasing \\ 
$P_p$ &$P_p$ & ferrule-GRIN lens& first prism surface&  $P_p$  &power \\ 
\hhline{|=|=|=|=|=|=|}
30 &21.6 &14.5 &11.01 &2.97 &0.017 \\ 
60 &42.6 &28.5 &22.03 &5.84 &0.032 \\ 
90 &62.8 &42.2 &33.04 &8.65 &0.044 \\
120 &83 &56 &44.06 &11.48 &0.057 \\ 
150 &103.1 &69.5 &55.07 &14.24 &0.195 \\ 
180 &123 &82.9 &66.09 &16.99 &0.37 \\ 
210 &142.3 &95.5 &77.10 &19.57 &0.63 \\
240 &161 &108.2 &88.12 &22.18 &1.33 \\
270 &179 &120.5 &99.13 &24.70 &2.32 \\
300 &197 &132.9 &110.15 &27.24 &3.21 \\
\hline
\end{tabular}
\end{center}
\end{table}

\section{Range of the Ti:sapphire WGL}
\begin{figure}[hbt!]
\centering
\includegraphics[width=0.9\linewidth]{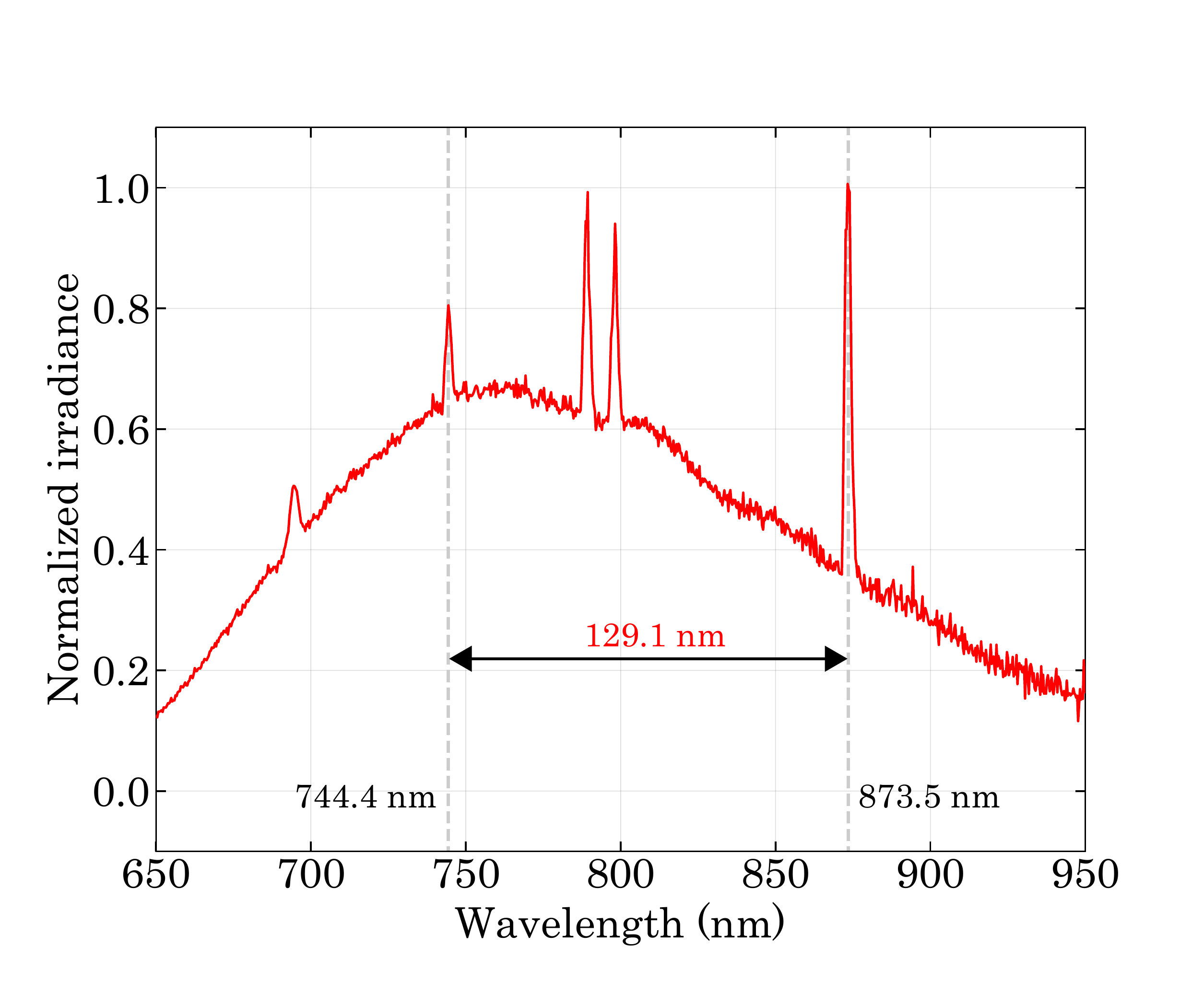}
\caption{The two furthest apart lasing peaks observed at \SI{744.4}{\nano\meter} and \SI{873.5}{\nano\meter}. This means that peaks are measured around a range of \SI{129.1}{\nano\meter}.}\label{fig:range}
\end{figure}

As shown in ~\ref{fig:range}, the Ti:sapphire WGL has a range of \SI{129.1}{\nano\meter}. The furthest two lasing peaks recorded are \SI{744.4}{\nano\meter} and \SI{873.5}{\nano\meter}.

\section{Exponential fits applied to individual linewidth sweeps}
\begin{figure}[hbt!]
\centering
\includegraphics[width=0.9\linewidth]{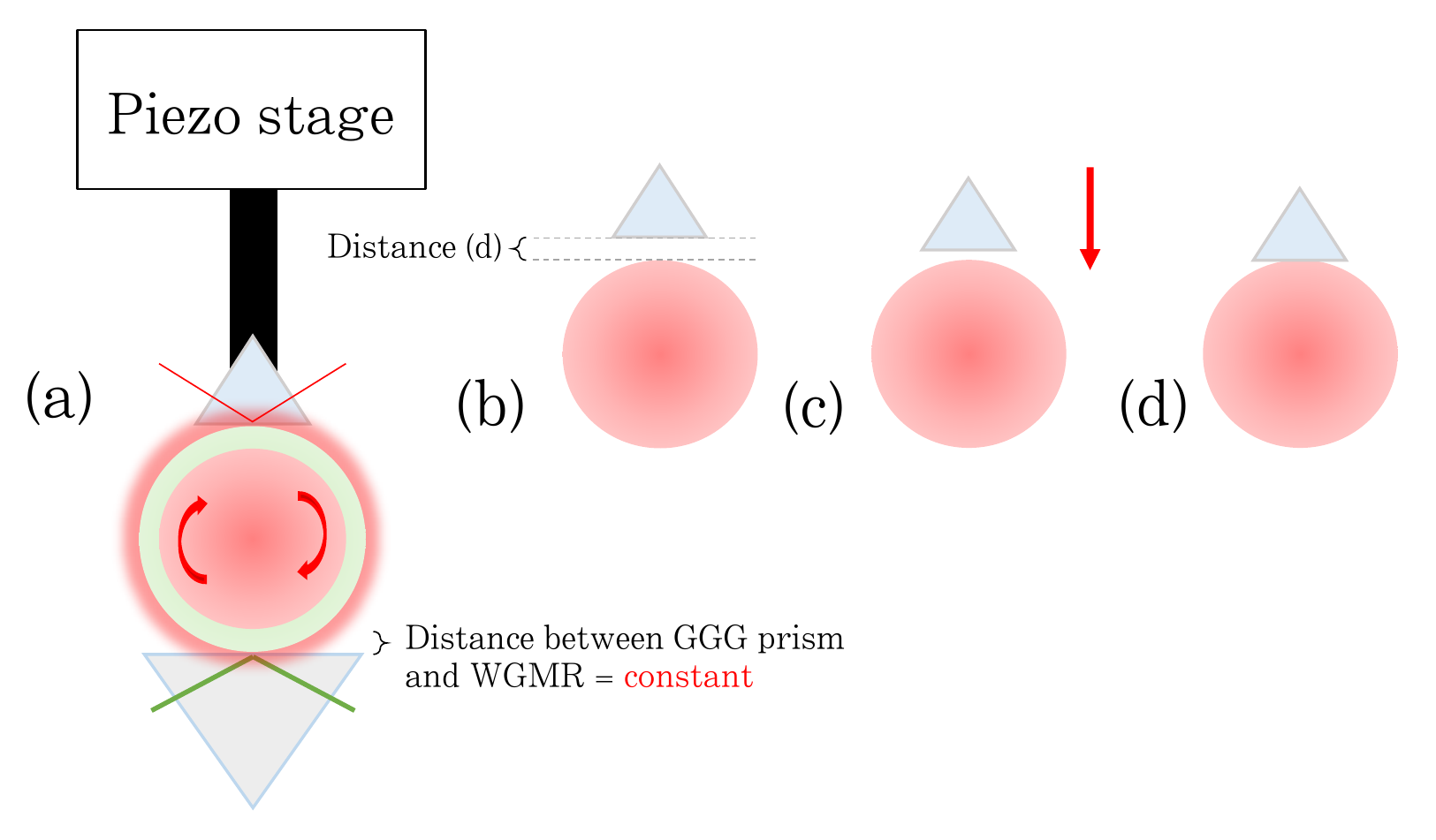}
\caption{Coupling distances of the GGG prism used to couple the pump and the diamond prism used to couple the probe laser to the Ti:sapphire WGMR. (a) shows a depiction of the image of the prisms which can be seen in Figure~\ref{fig:Gainssetupimage}(b). (b) shows the distance (d) between the resonator and the prism, this distance is varied by using the piezo stage shown in (a) and is plotted on each primary axis of Figure~\ref{fig:LWsweep}. (c) shows that the prism is being moved towards the resonator, i.e., d is being decreased. (d) shows that the prism and WGMR are in contact, this can be observed in each sub-figure of Figure~\ref{fig:LWsweep} in the form of a dotted line.}\label{fig:couplin795}
\end{figure}

\begin{figure}[t!]
\centering
\includegraphics[width=0.9\linewidth]{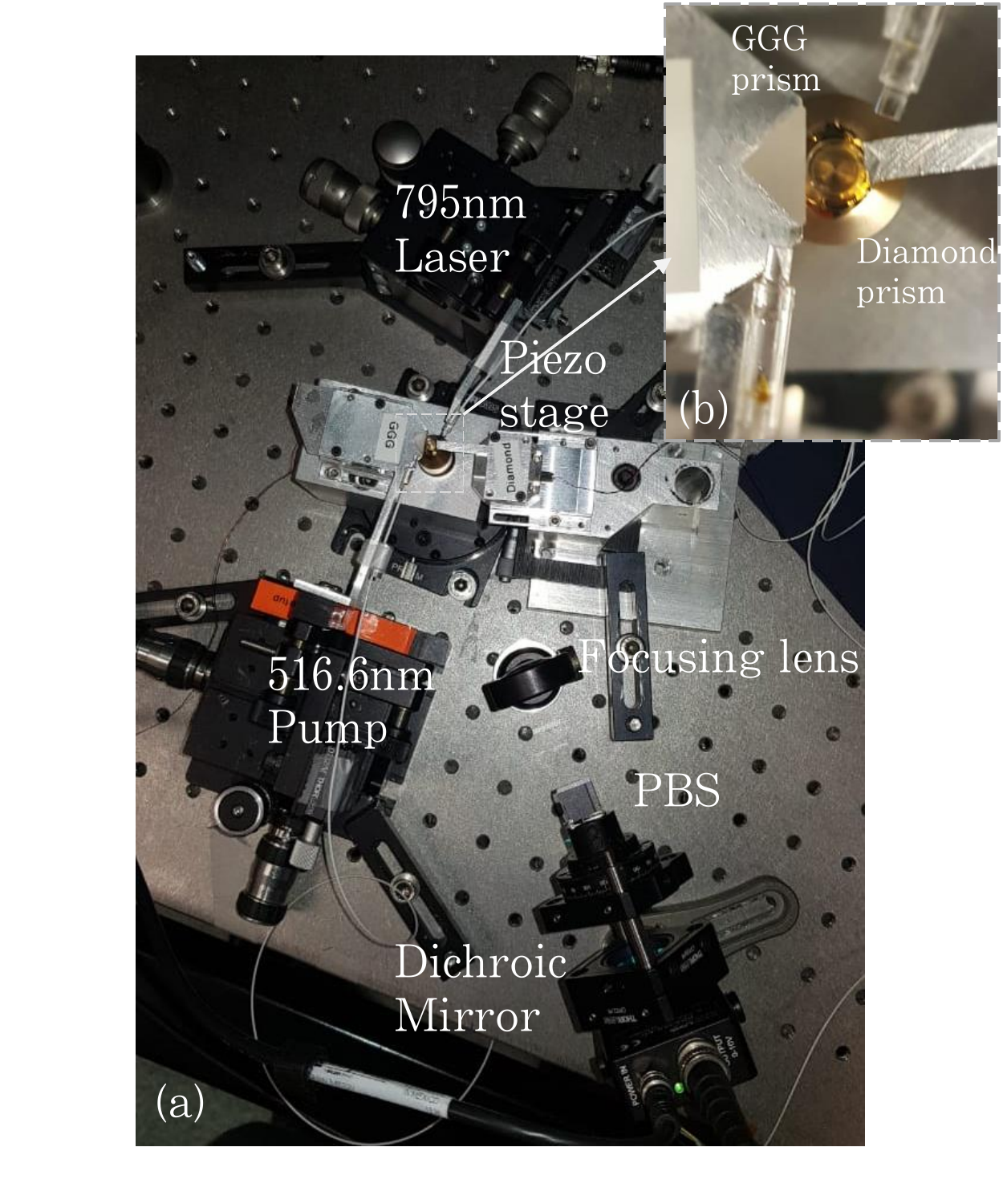}
\caption{Image of the modified experimental setup used to observe amplification with the \SI{795}{\nano\m} probe laser. (a) shows majority of the experimental components. (b) shows a zoomed in image of the two prisms used to couple the gain and the probe lasers to the Ti:sapphire WGMR. The pigtailed ferrules and GRIN lenses are also visible in the image. The function of the piezo stage is shown in Figure~\ref{fig:couplin795}.}\label{fig:Gainssetupimage}
\end{figure}

\begin{figure}
  \centering
  \includegraphics[width=0.32\linewidth]{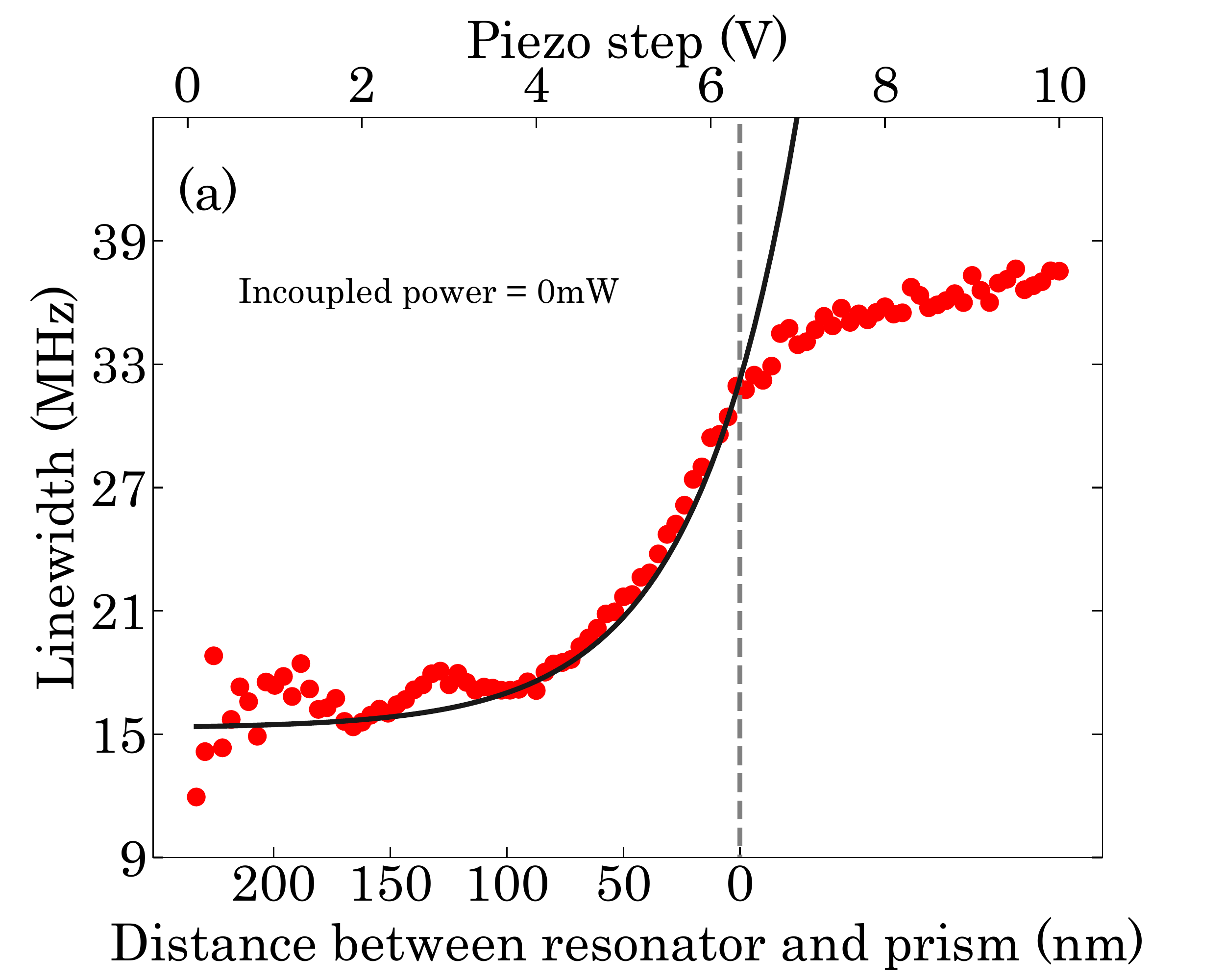}
  \includegraphics[width=0.32\linewidth]{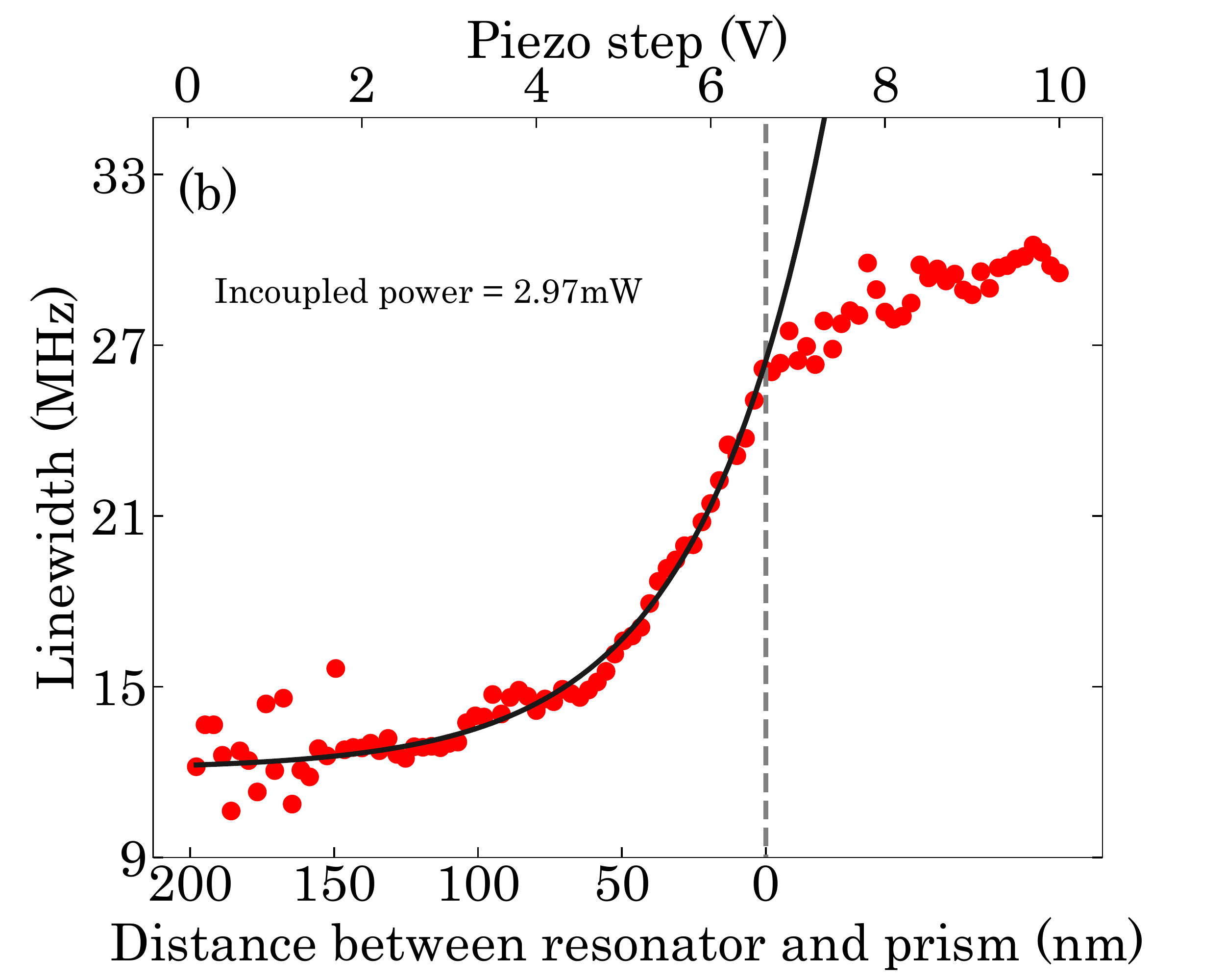}
  \includegraphics[width=0.32\linewidth]{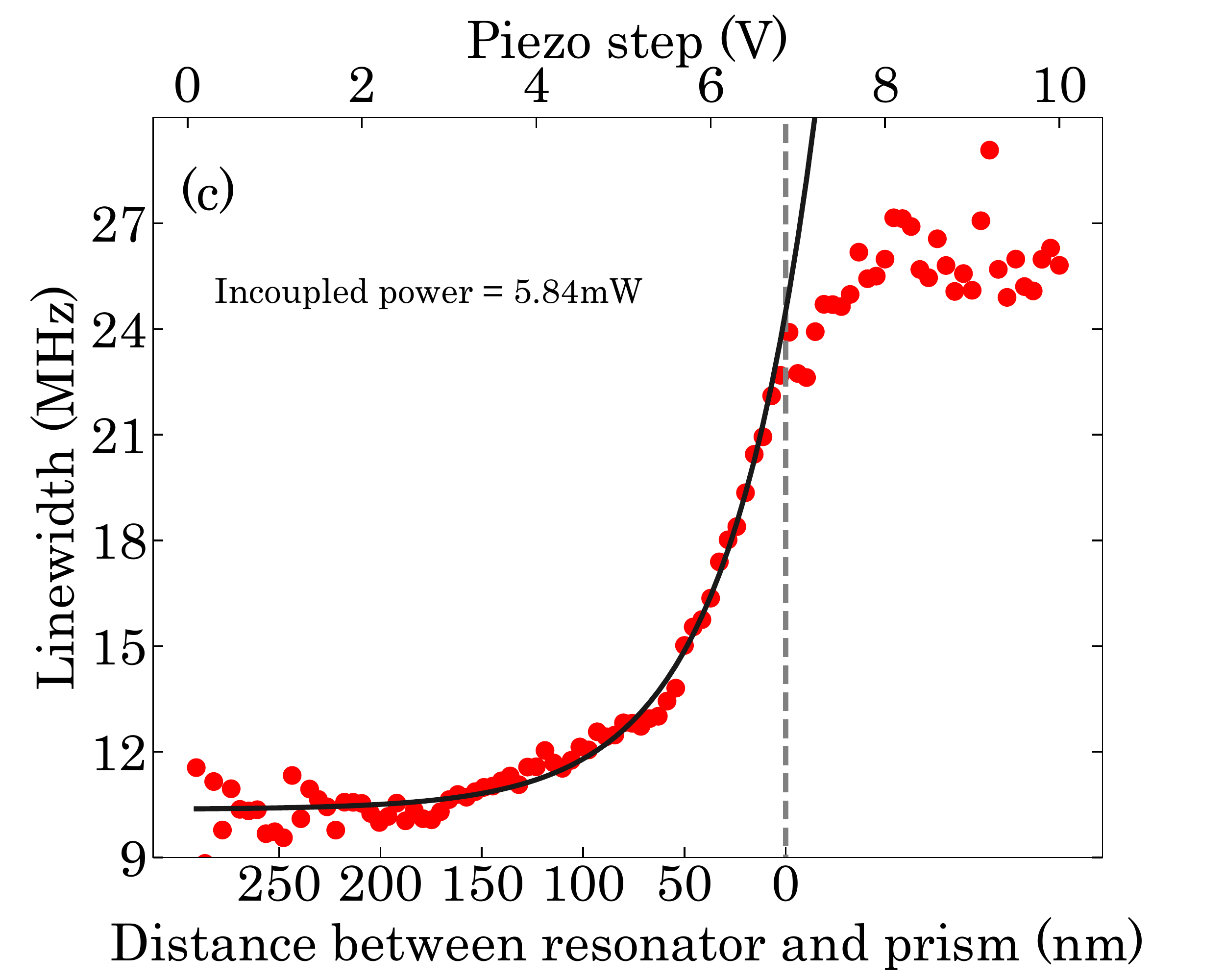}\\
  \includegraphics[width=0.32\linewidth]{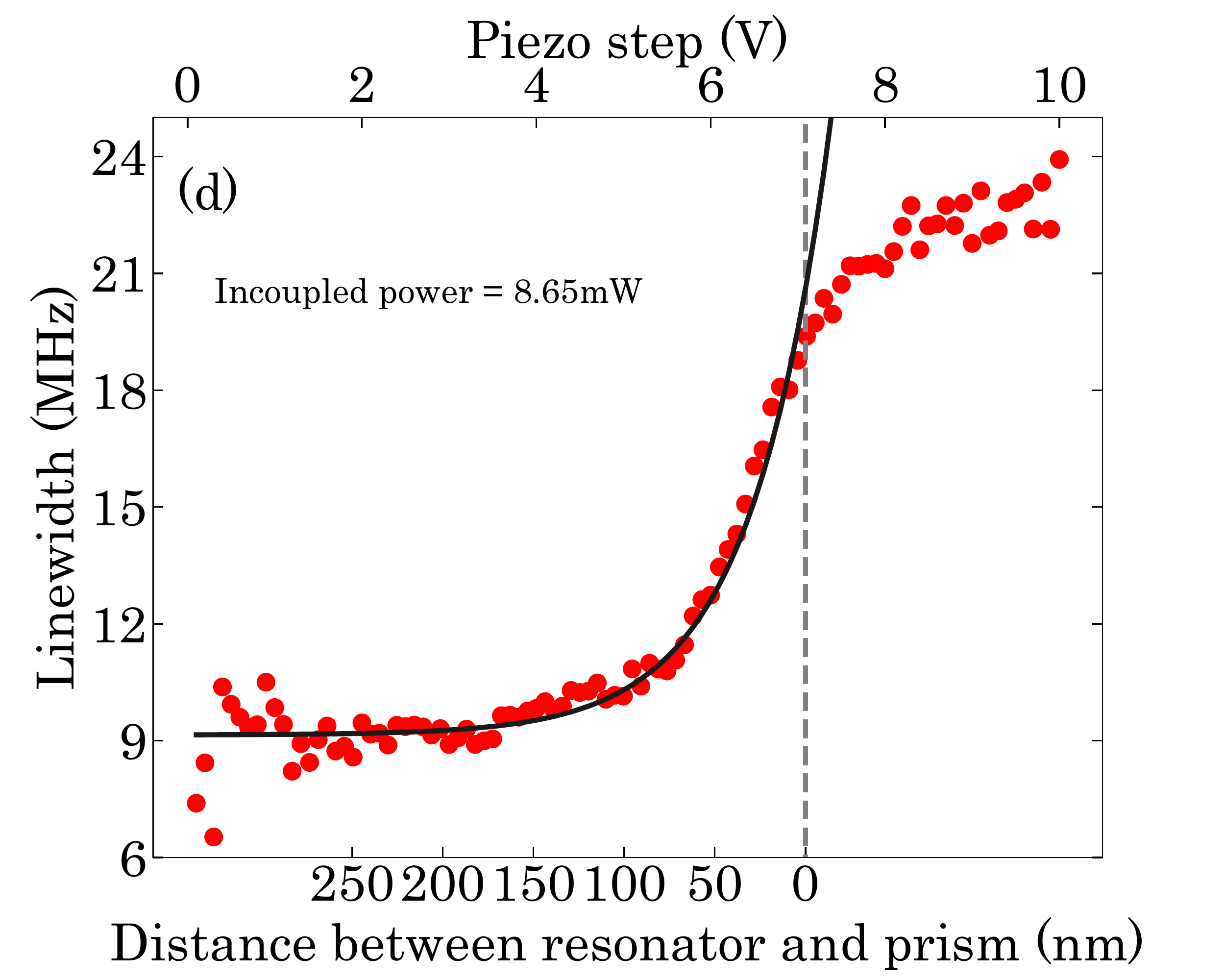}
  \includegraphics[width=0.32\linewidth]{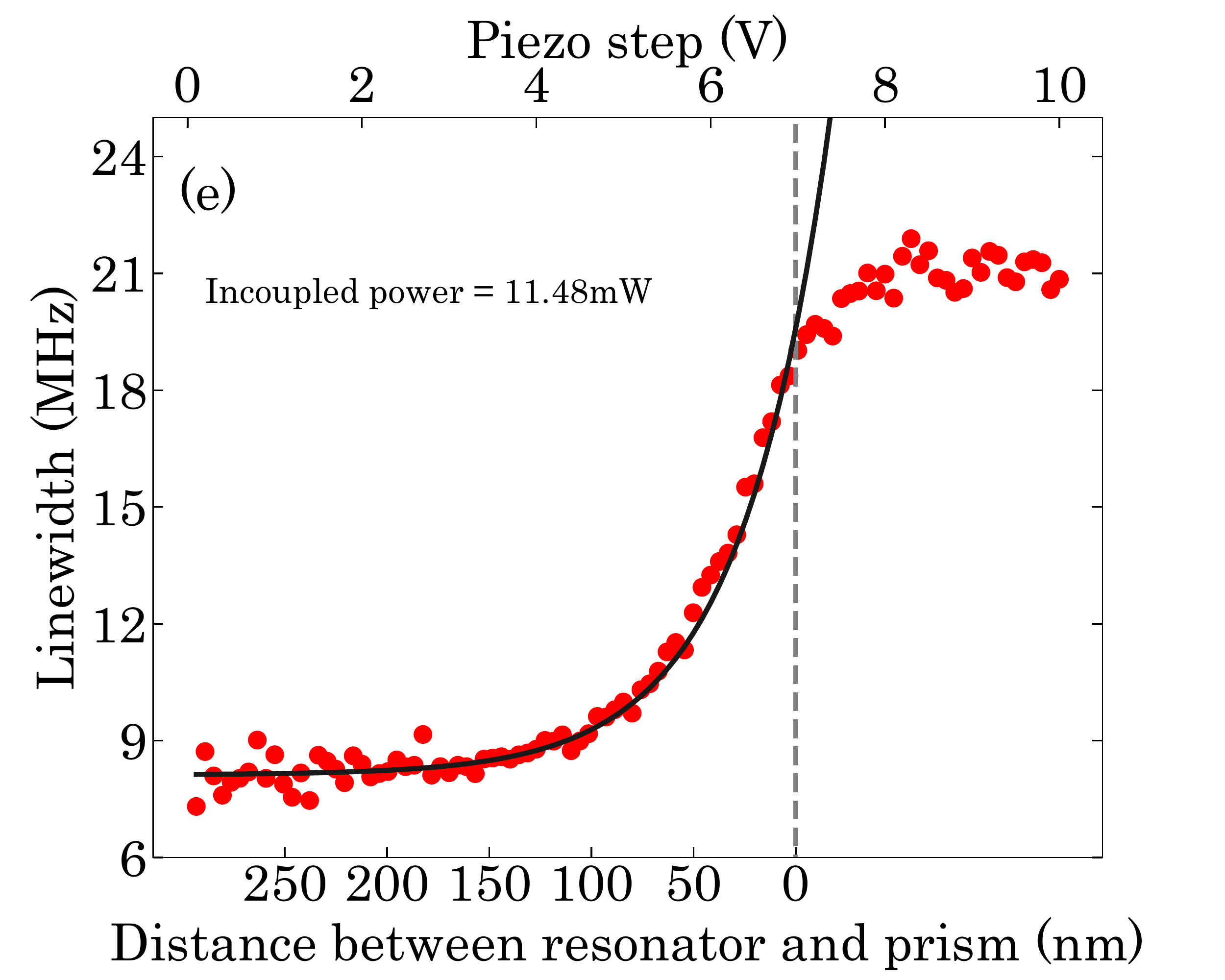}
  \includegraphics[width=0.32\linewidth]{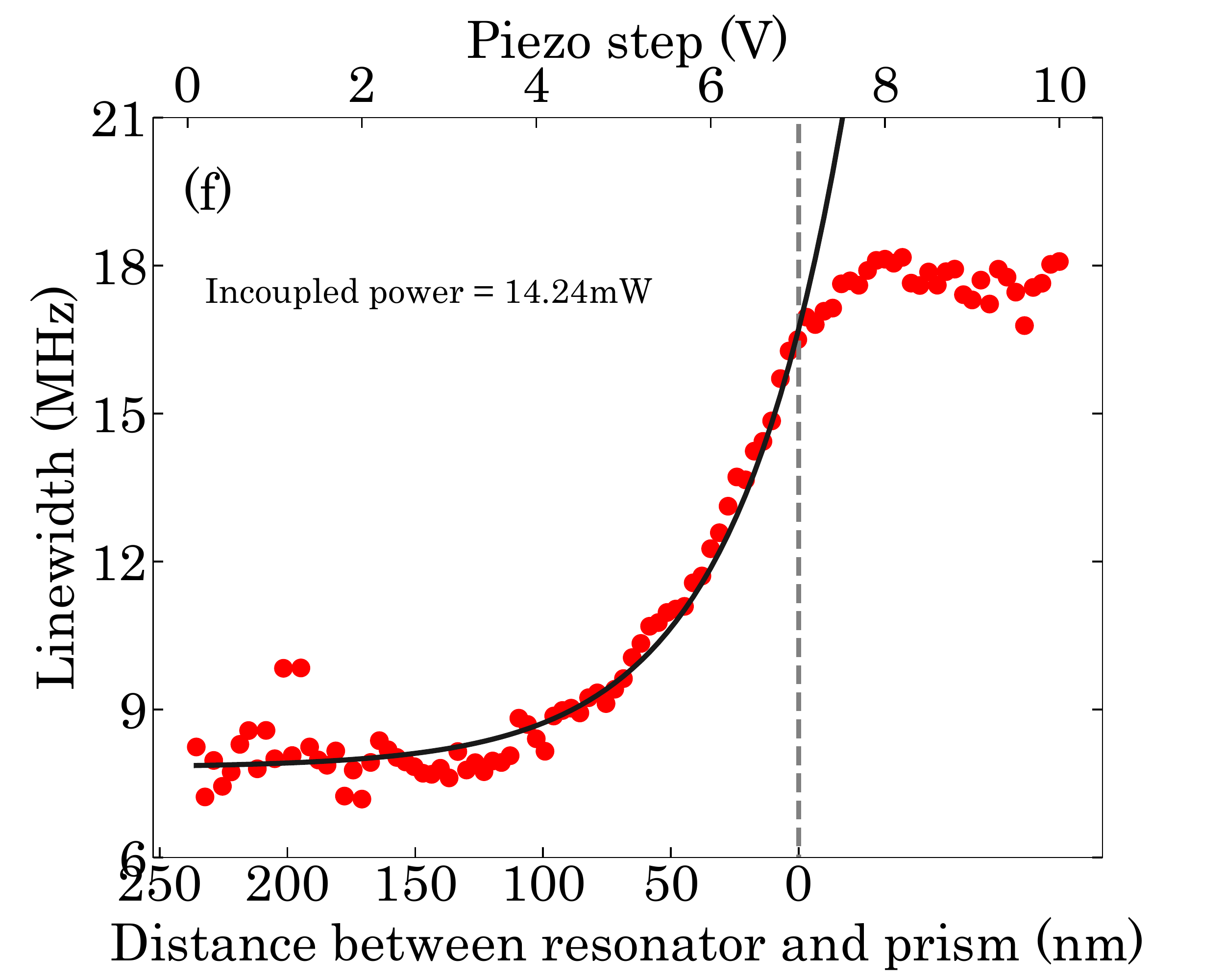}\\
  \includegraphics[width=0.32\linewidth]{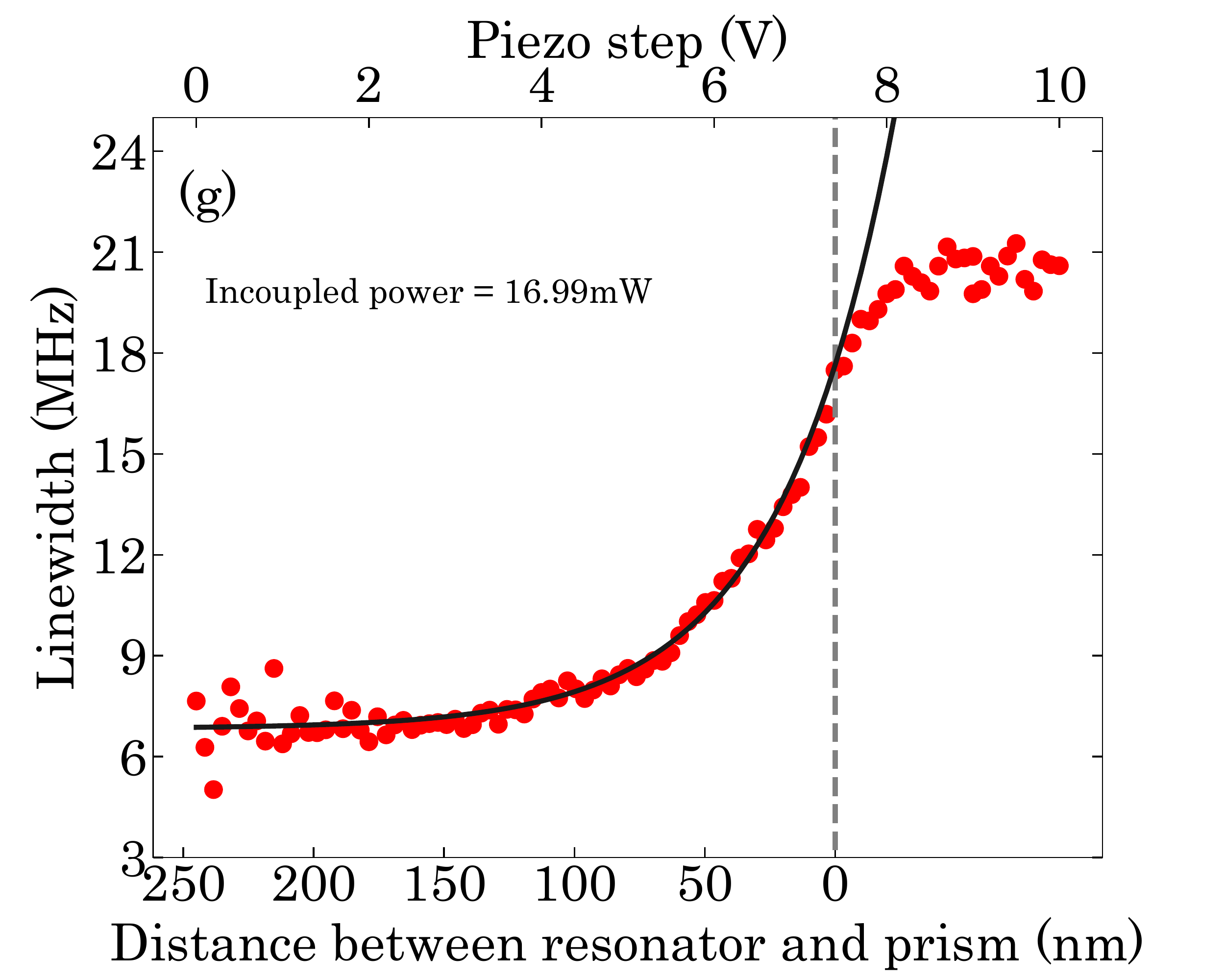}
  \includegraphics[width=0.32\linewidth]{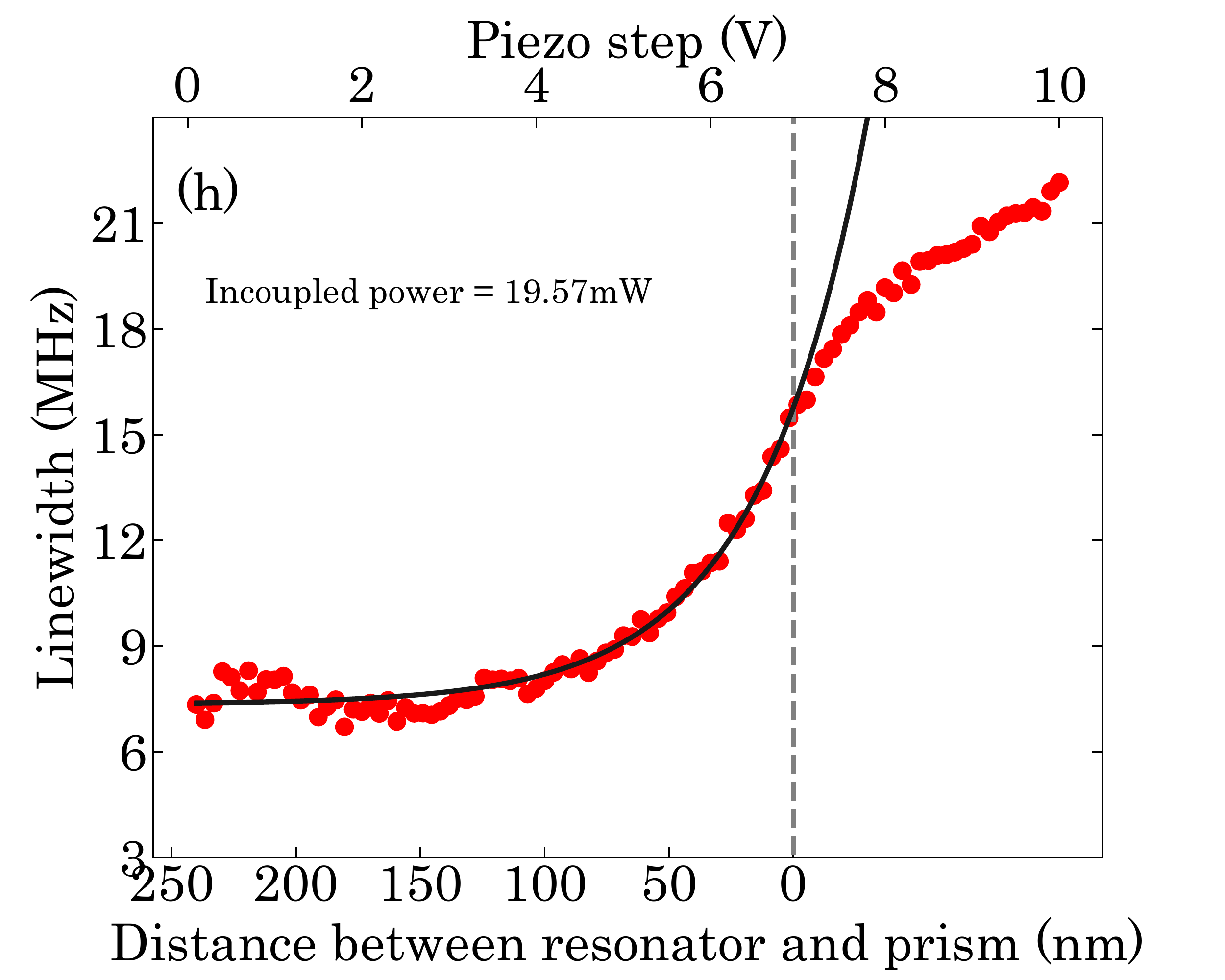}
  \includegraphics[width=0.32\linewidth]{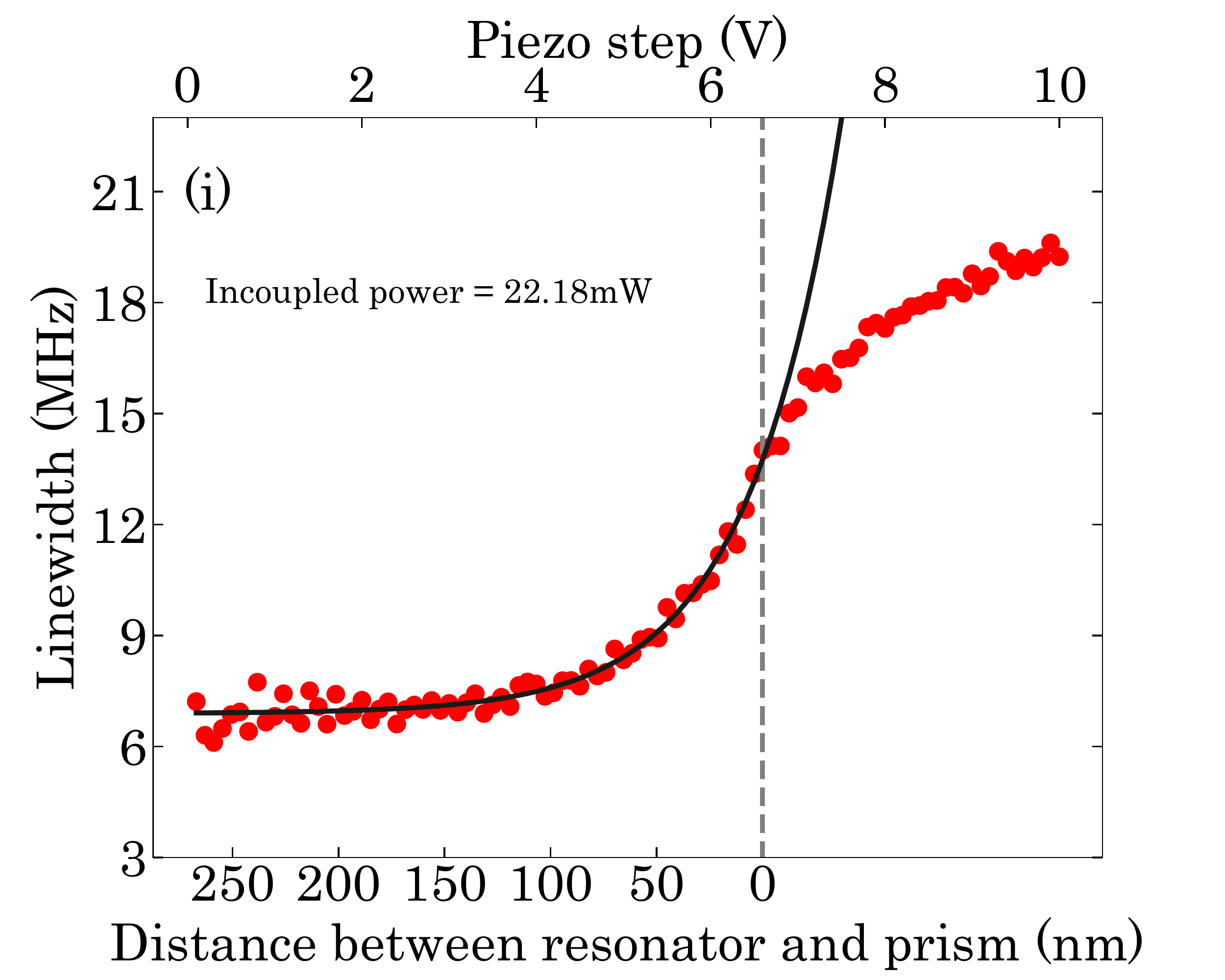}\\
  \includegraphics[width=0.32\linewidth]{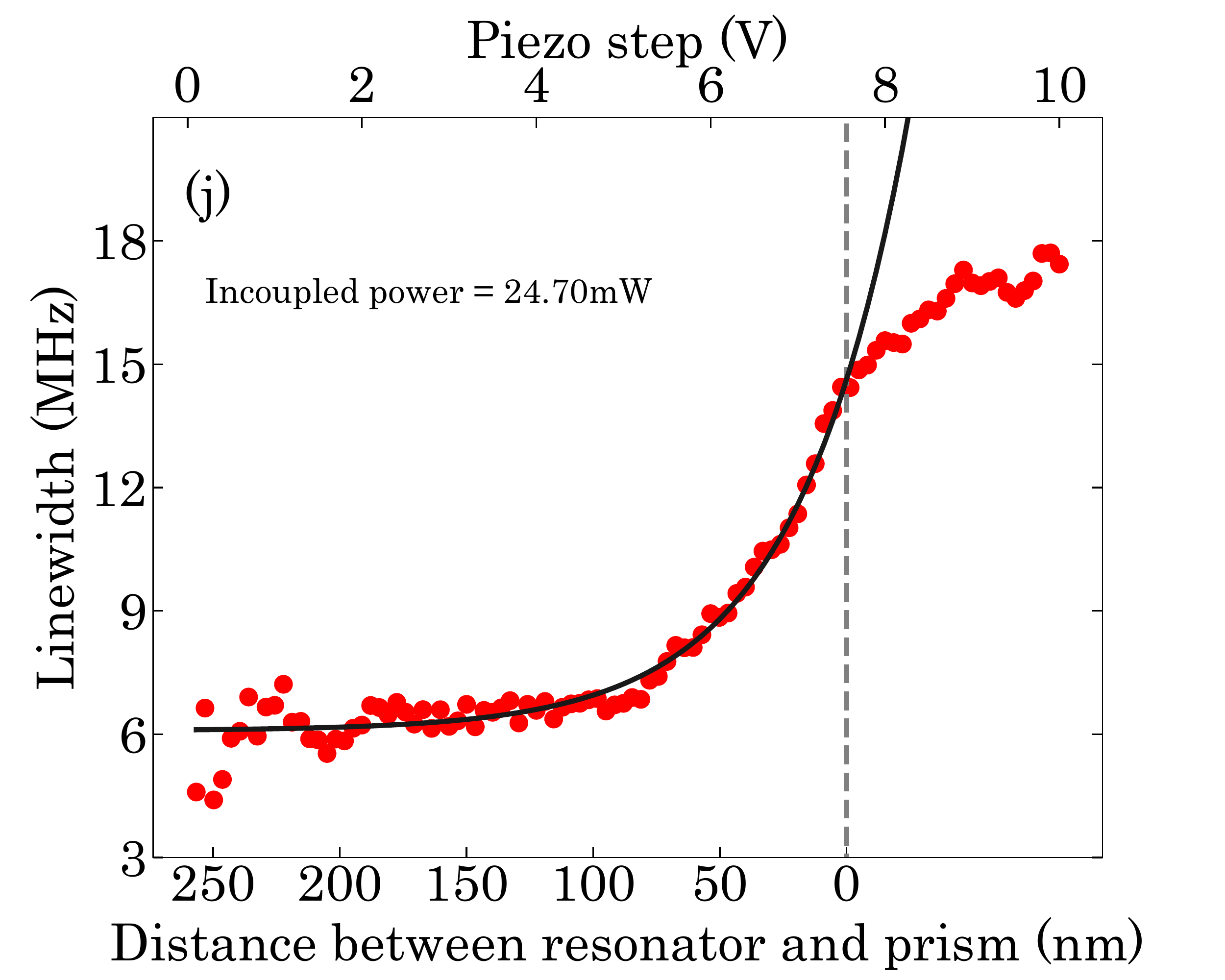}
  \includegraphics[width=0.32\linewidth]{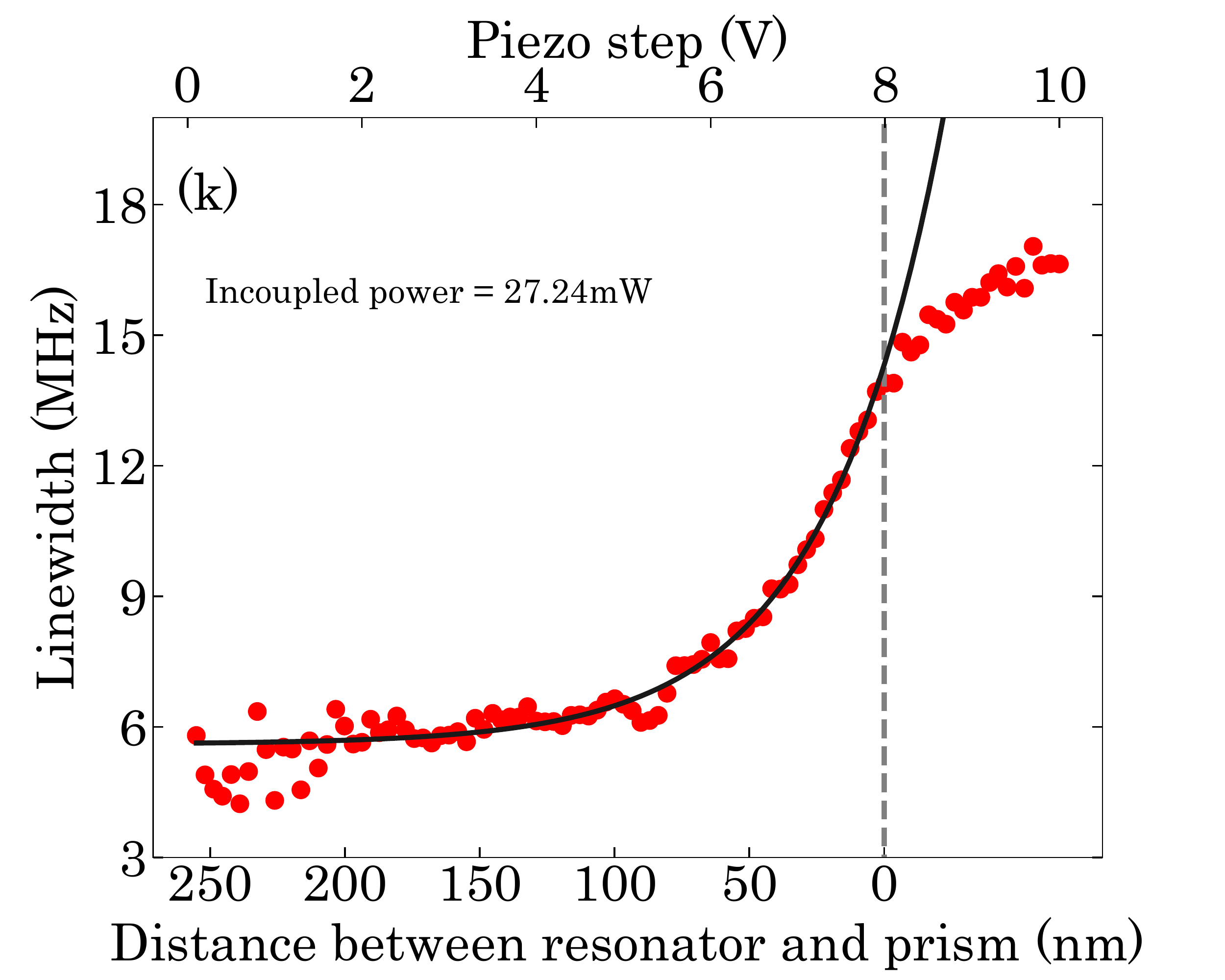}
  \caption{Various linewidths of different pump powers used to create Figure 7 in the main text. The experimental data is shown in red and the theoretical exponential fits are shown in black, (a) gives the probe WGM linewidth at \SI{0}{\milli\watt}, i.e., when the pump is off, (b) gives the linewidth of the probe WGM at an incoupled pump power of \SI{2.97}{\milli\watt}. Sub-figures (c) to (k) are plotted for pump powers listed in the second last column of table~\ref{tab:PumpPower} from right, i.e., incoupled $P_p$ in , in ascending order of power with equal increments. the dotted line shows the \SI{0}{\nano\meter} distance between the WGMR and the prism, i.e., that they are in contact. It can be noted that beyond this point the data points do not follow the exponential trend, i.e., the exponential does not fit to the data. Due to this reason in Figure 7 of the main the exponential fits are only plotted up to the dotted line.
  }
  \label{fig:LWsweep}
\end{figure}

 In order to record the linewidth data shown in Figure 7(a) (main text), a piezo stage was used to move the diamond prism prism towards or away from the WGMR. This configuration is shown in Figure~\ref{fig:couplin795}. The distance between the GGG prism used to couple the pump laser was kept constant whereas the piezo was used to change the distance between the piezo and the resonator.  Each sub-figure of Figure~\ref{fig:LWsweep} has a primary and a secondary horizontal axis with the primary horizontal axis displaying the distance between the resonator and the prism, which was calculated using the piezo voltage displayed by the secondary horizontal axis. At each iteration of the pump laser, a high $Q$ mode excited using the \SI{795}{\nano\m} probe laser was monitored, i.e., the linewidth of the mode was recorded at different coupling rates, i.e., under-coupled to over-coupled or in other words with the prism far away to prism touching the resonator, as depicted in Figure~\ref{fig:couplin795}. An image of the setup for the amplification experiment is shown in Figure~\ref{fig:Gainssetupimage}. The following equation was used to apply the exponential fits to the data shown in Figure 7 (main text):

\begin{equation}
    y = a + b e^{cx}
\end{equation}
coefficients $a$, $b$ and $c$ were extracted for each fit and then used to find the distance ($d$) between the resonator and the prism (shown in Figure~\ref{fig:couplin795}(b)) using the following equation:

\begin{equation}
    d = -\frac{c}{2\kappa}(\mathrm{V} - \mathrm{V_\text{touch}})
\end{equation}
where V is the piezo voltage at each individual step whereas V$_\text{touch}$ is the voltage at which the prism and the resonator come into contact. The evanescent-field decay constant ($\kappa$) is given by~\cite{gorodetsky1999optical,trainor2018selective}: 
\begin{equation}
    \kappa = \frac{2\pi}{\lambda}\sqrt{n_\text{res}^2 - n_\text{out}^2}
\end{equation}
where $n_\text{res}$ is the refractive index of the resonator and $n_\text{out}$ is the refractive index of the surrounding medium of the resonator, i.e., air. $V_\text{touch}$ is denoted in each sub-figure of Figure~\ref{fig:LWsweep} by the dotted line. It can be observed that the data does not follow an exponential trend after the dotted line. In order to record the data shown in Figure~\ref{fig:LWsweep}, a piezo stage was utilized to move the diamond coupling prism towards the resonator

Figure~\ref{fig:LWsweep} shows the individual exponential fits applied to the linewidth sweep data acquired in the amplification experiment. Figure~\ref{fig:LWsweep}(a) shows the linewidth data and the exponential fit (in black) when the pump laser is off, Figure~\ref{fig:LWsweep}(b) shows the data when the pump is turned on and \SI{2.97}{\milli\W} of power is coupled into the resonator, Figure~\ref{fig:LWsweep}(c) shows the data when \SI{5.84}{\milli\W} of power is coupled into the resonator and vice versa, ending at Figure~\ref{fig:LWsweep}(k) which shows the data when \SI{27.24}{\milli\W} of power is coupled into the resonator.

\clearpage